\RequirePackage{xcolor}
\RequirePackage{ifpdf}
\documentclass[letterpaper]{JHEP3}
\usepackage{amsmath}
\usepackage{epsfig}
\usepackage{subfigure}
\usepackage{enumitem}

\pdfoutput=1

\newcommand{\roughly}[1]{\mathrel{\raise.3ex\hbox{$#1$\kern-0.85em
\lower1ex\hbox{$\sim$}}}}

\def\nn{\nonumber}

\newcommand{\be}{\begin{equation}}
\newcommand{\bee}{\begin{equation}}
\newcommand{\ee}{\end{equation}}
\newcommand{\beea}{\begin{eqnarray}}
\newcommand{\eea}{\end{eqnarray}}
\newcommand{\bea}{\begin{eqnarray}}

\def\nott#1{\setbox0=\hbox{$#1$}                
   \dimen0=\wd0                                 
   \setbox1=\hbox{/} \dimen1=\wd1               
   \ifdim\dimen0>\dimen1                        
      \rlap{\hbox to \dimen0{\hfil/\hfil}}      
      #1                                        
   \else                                        
      \rlap{\hbox to \dimen1{\hfil$#1$\hfil}}   
      /                                         
   \fi}                                         %
\def\Dsl{\nott{D}}

\def\uxsl{\hbox{/\kern-.4000em$u$}}
\def\uxslsm{\hbox{\smaller/\kern-.5600em$u$}}
\def\pxpsl{\hbox{/\kern-.5000em$p$}}
\def\epssl{\hbox{/\kern-.5600em$\epsilon$}}
\def\delsl{\hbox{/\kern-.7000em$\nabla$}}
\def\lxpsl{\hbox{/\kern-.5600em$l$}}
\def\kxpsl{\hbox{/\kern-.5600em$k$}}
\def\qxpsl{\hbox{/\kern-.3900em$q$}}

\def\pref#1{(\ref{#1})}
\def\exd{{\rm d}}
\def\ol#1{{\overline{#1}}}

\def\cB{{\cal B}}
\def\cD{{\cal D}}

\def\cE{{\cal E}}
\def\cF{{\cal F}}

\def\cH{{\cal H}}

\def\cL{{\cal L}}

\def\cN{{\cal N}}

\def\cR{{\cal R}}

\def\cU{{\cal U}}

\def\cW{{\cal W}}
\def\cX{{\cal X}}
\def\cY{{\cal Y}}

\def\mfa{{\mathfrak a}}
\def\mfb{{\mathfrak b}}
\def\mfc{{\mathfrak c}}

\def\mfe{{\mathfrak e}}
\def\mff{{\mathfrak f}}
\def\mfg{{\mathfrak g}}
\def\mfh{{\mathfrak h}}

\def\mfm{{\mathfrak m}}

\def\mfv{{\mathfrak v}}
\def\mfw{{\mathfrak w}}

\def\mfM{{\mathfrak M}}
\def\mfP{{\mathfrak P}}

\def\mfZ{{\mathfrak Z}}

\def\ssB{{\scriptscriptstyle B}}

\def\ssD{{\scriptscriptstyle D}}
\def\ssE{{\scriptscriptstyle E}}
\def\ssF{{\scriptscriptstyle F}}

\def\ssH{{\scriptscriptstyle H}}
\def\ssI{{\scriptscriptstyle I}}

\def\ssL{{\scriptscriptstyle L}}

\def\ssR{{\scriptscriptstyle R}}

\def\ssU{{\scriptscriptstyle U}}

\def\ssW{{\scriptscriptstyle W}}
\def\ssX{{\scriptscriptstyle X}}
\def\ssY{{\scriptscriptstyle Y}}

\def\UV{{\scriptscriptstyle UV}}

\def\SM{{\scriptscriptstyle SM}}

\def\IR{{\scriptscriptstyle IR}}
\def\SG{{\scriptscriptstyle SG}}

\title{Who's Afraid of the Supersymmetric Dark?\\The Standard Model vs Low-Energy Supergravity
}

\author{C.P.~Burgess,${}^{1,2,3}$ and F.~Quevedo${}^{4}$\\

{\it 
${}^1$ Department of Physics \& Astronomy, McMaster University, 
 Hamilton ON, Canada.\\
${}^2$ Perimeter Institute for Theoretical Physics, 
Waterloo ON, Canada.\\
${}^3$ CERN, Theoretical Physics Department, Gen\`eve 23, Switzerland.\\
${}^4$ DAMTP, Cambridge University, Wilberforce Road,  Cambridge, CB3 0WA, UK.
}
}

\date{\today}

\abstract{Use of supergravity equations in astronomy and late-universe cosmology is often criticized on three grounds: ($i$) phenomenological success usually depends on the supergravity form for the scalar potential applying at the relevant energies; ($ii$) the low-energy scalar potential is extremely sensitive to quantum effects involving very massive particles and so is rarely well-approximated by classical calculations of its form; and ($iii$) almost all Standard Model particles count as massive for these purposes and none of these are supersymmetric. Why should Standard Model loops preserve the low-energy supergravity form even if supersymmetry is valid at energies well above the electroweak scale? We use recently developed tools for coupling supergravity to non-supersymmetric matter to estimate the loop effects of heavy non-supersymmetric particles on the low-energy effective action, and provide evidence that the supergravity form is stable against integrating out such particles (and so argues against the above objection). This suggests an intrinsically supersymmetric picture of Nature where supersymmetry survives to low energies within the gravity sector but not the visible sector (for which supersymmetry is instead non-linearly realized). We explore the couplings of both sectors in this picture and find that the presence of auxiliary fields in the gravity sector makes the visible sector share many features usually attributed to linearly realized supersymmetry although (unlike for the MSSM) a second Higgs doublet is not required for all Yukawa couplings to be non-vanishing and changes the dimension of the operator generating the Higgs mass. We discuss the naturalness of this picture and some of the implications it might have when searching for dark-sector physics.}

\begin{document}

\section{Introduction}

The absence of evidence for superpartners at the Large Hadron Collider (LHC) \cite{ATLAS:2015wrn,Sarkar:2021lju, ParticleDataGroup:2020ssz}
makes supersymmetry as a solution to the hierarchy problem appear to be a beautiful idea mugged by a gang of ugly facts. Yet supersymmetry remains well-motivated at very high energies; appearing to play a central role there in frameworks like string theory \cite{Green:2012oqa,
Polchinski:1998rr, Becker:2006dvp,Ibanez:2012zz} that sensibly quantum-complete gravity at the highest scales. 

But as the hierarchy problem recedes as a motivation, seeking supersymmetry in accelerators is like searching under the proverbial streetlight on a dark night; absence of success might be more about the search strategy than indicating that searches are a fruitless exercise. Putting ease of detection aside, is there a place we should expect supersymmetry is most likely to arise (and so be the most relevant for understanding) if it exists? 

One way to approach this question is to ask: for which kinds of particles should we expect the mass splittings between superpartners to be the smallest? In supersymmetric theories supersymmetry is usually broken when some field $F$ acquires an expectation value. The size of the mass splittings between bosons and fermions within any particular multiplet are then in order of magnitude given by 
\be
    m_\ssB^2 - m_\ssF^2 \sim g F \,,
\ee 
where $g$ is a measure of the strength of interaction between $F$ and the multiplet of interest. This suggests that the particles that are split the least are also those that couple the weakest. 

Gravity is the weakest known interaction and there are even circumstantial reasons to believe that it might be the weakest interaction that is possible \cite{Arkani-Hamed:2006emk}. If so, then it is natural to expect that it is the gravitational sector that should be the most supersymmetric, and this is indeed what often happens in explicit higher-dimensional supergravity models \cite{SLED, HighEUV}. Perhaps the gravity sector is the one where the implications of low-energy supersymmetry are most prominent. If so, the very weakness of gravitational interactions helps make supersymmetry's detection more difficult. 

Despite its weakness we certainly know that gravity exists and its properties are measured in great detail within the solar system, in astronomy and in cosmology. The above line of argument suggests that these areas might also be among the best venues for seeking evidence for supersymmetry of this type at low energies. It is perhaps unsurprising from this point of view that current evidence for dark sectors dominantly comes from these kinds of observations. It also has motivated studies \cite{SUGRAQuint, Binetruy:2009zz} that explore the implications of supersymmetric models at the lowest possible energies, like those relevant to late-time cosmology (often as variants of quintessence models  \cite{quintessence, QuintessenceReviews} for dark energy).

\subsubsection*{An objection}

That paints a pleasing picture, but there is a long-standing objection to using supergravity in late-time cosmology in this way. Success or failure in cosmological models often turns on the detailed properties of the scalar potential, and many of the ingredients required for success in cosmology ({\it e.g.}~extremely light scalars and small vacuum energies) are known to be exquisitely sensitive to quantum effects involving the theory's highest-energy sector \cite{CCWeinberg, Burgess:2013ara}. Ordinary Standard Model particles (like the electron) count as high-energy particles from the point of view of cosmology, and these are measured not to be supersymmetric at all. The objection asks how the putative supersymmetry of scalar potentials relevant to the extremely low energies of cosmology could possibly survive the quantum corrections generated by integrating out the known non-supersymmetric Standard Model particles. 

A difficulty in making this objection definitive has been the inability to precisely formulate a theory in which a very supersymmetric gravity sector consistently couples to non-supersymmetric matter. Precisely formulating this type of theory is a prerequisite for computing quantum corrections to the low-energy scalar potential, and only these kinds of calculations can rule out or verify the prejudice that quantum corrections involving the known particle spectrum should ruin the supersymmetry of the low-energy world of cosmology.

The purpose of this paper is to re-examine this issue in view of recent progress understanding how to couple supergravity to non-supersymmetric matter \cite{Komargodski:2009rz, Bergshoeff:2015tra, Dudas:2015eha, DallAgata:2015zxp, Schillo:2015ssx}. In \S\ref{sec:SuperNonlin} we review this new understanding and in \S\ref{ssec:Naturalness} we use it to estimate the size of the quantum corrections that arise once heavy non-supersymmetric particles are integrated out. We are able to draw the following conclusions.
\begin{itemize}
\item{We describe a concrete scenario in which supersymmetry is linearly realized in a gravitational (hidden) sector and non-linearly realized in the Standard Model (visible) sector. Constrained superfields are used to describe the Standard Model fields. Contrary to the MSSM, there is no need from anomaly cancellation to introduce a second Higgs superfield (nor  a $\mu$-term, since the higgsino is integrated out). Couplings of the goldstino superfield to the Standard Model sector determine the Higgs potential and the up-quark Yukawa couplings.}

\item We find no evidence that loops of non-supersymmetric Standard Model particles must destabilize the general form of the Lagrangian used to couple supergravity to non-supersymmetric matter, supporting the consistency of using supergravity for late-time cosmology. 
\item Supersymmetry does not in itself automatically solve the questions of technical naturalness that arise in low-energy cosmological applications (such as the cosmological constant problem or the tuning problems of quintessence theories), making it necessary to check these on a model-by-model basis. It is interesting though that the interplay between gravity and the size of the scalar potential plays a central role in most of these problems, so having a supersymmetric gravity sector could plausibly be part of the final picture that resolves them.\footnote{See \cite{CCNo-Scale} for an approach that combines this low-energy supersymmetric framework with the general scale-invariance arguments of \cite{Burgess:2020qsc} to address low-energy naturalness problems in quintessence models.} 
\item The low-energy scalar potential of these theories closely resembles the usual supergravity structure. This structure remains stable as non-supersymmetric particles are integrated out because of the presence of the auxiliary fields associated with the gravity, goldstino (and possibly other) supermultiplets that (by assumption) appear in the low-energy theory due to the assumed supersymmetry of the gravity sector. Although these auxiliary fields do not propagate they do affect how loops of heavy particles contribute to the low-energy theory, making their inclusion crucial for understanding naturalness issues.\footnote{The importance of auxiliary fields resembles the important role played by non-propagating `topological' fields in other areas of physics like the Quantum Hall Effect \cite{QHE, EFTBook}. This connection is strengthed by the observation that auxiliary fields arise as 4-form fields in string theory \cite{LuisForm, Burgess:2020qsc}, with 4-forms known to bring information about higher-dimensional topology into the low-energy 4D theory, both in string theory and more broadly in extra-dimensional models \cite{MyForm}.}
\item Large positive vacuum energies are relatively common features of models coupling supergravity to non-supersymmetric matter and this might indicate that this framework is also useful for understanding inflationary models of the much earlier universe (as has indeed been explored in \cite{Inflation}). 
\end{itemize} 
 
\subsubsection*{Gravity's dark side}

A universe in which non-supersymmetric Standard Model particles couple to a very supersymmetric gravitationally interacting dark sector obviously provides both observational challenges and phenomenological opportunities. Although there is some freedom choosing the dark sector's particle content, the structure imposed by supersymmetry also carries many constraints. Accidental scale invariance often ensures the existence of at least one axio-dilaton supermultiplet, whose spin-half superpartner (the dilatino) has a mass similar to the gravitino. One or more such multiplets are common in theories describing the low-energy limit of higher dimensional models due to the generic role played in them by scale invariance \cite{Burgess:2020qsc}. 

Besides the compulsory existence of a relatively light gravitino this picture generically contains many dark-matter candidates, including axions or other massive particles coupling to the Standard Model only very weakly. This provides a natural origin for low-energy axions, such as in `axiverse' models \cite{axiverse}, as well as Planck-coupled dark matter models \cite{Garny:2015sjg}. Although gravitationally coupled fermions would be impossible to see in underground detectors, they become sterile neutrinos (and so could become detectable) once they mix with Standard Model neutrinos, providing a potentially rich source of phenomenology \cite{Dienes:1998sb} and potential mechanisms for baryo- and lepto-genesis. We briefly describe some of the potential phenomenological consequences in \S\ref{sec:Conclusions} below.

\section{Supergravity coupled to non-supersymmetric matter}
\label{sec:SuperNonlin}

This section describes how to consistently couple supersymmetric gravity to ordinary matter that is assumed not to be  supersymmetric at all. This is the effective theory that one would expect at electroweak energies in a world in which the Standard Model sector couples more strongly to the supersymmetry-breaking fields than does the gravity multiplet. For the present purposes it is the naturalness properties of this construction -- in particular the stability of the small splitting in the gravity multiplet -- that are of greatest interest.

There is no loss of generality in describing such systems using the formalism of nonlinearly realized supersymmetry as formulated by \cite{Komargodski:2009rz}, together with its coupling to supergravity \cite{Bergshoeff:2015tra, Dudas:2015eha, DallAgata:2015zxp, Schillo:2015ssx}. It is the generality of this construction that ultimately underlies its stability under Wilsonian evolution. 

The logic of this construction goes as follows. Reference \cite{Komargodski:2009rz} first shows how to take an {\it arbitrary} non-supersymmetric theory and rewrite it `as if' it were globally supersymmetric.\footnote{The same is also possible for more mundane symmetries: a generic non-invariant action can always be made invariant under a global symmetry by appropriately coupling to the relevant Goldstone bosons \cite{Weinberg:1968de, Coleman:1969sm, Callan:1969sn}.} This can always be done for global supersymmetry simply by coupling the non-supersymmetric matter appropriately to the Goldstone fermion \cite{VolkovAkulov} whose presence is always required in the low-energy sector of any system whose UV supersymmetry is spontaneously broken (as it must be if supermultiplets are split badly enough to allow some of its members to be integrated out while the others are not). The coupling of this system to supergravity -- described in \cite{Bergshoeff:2015tra, Dudas:2015eha, DallAgata:2015zxp, Schillo:2015ssx} -- then follows as a special case of the usual rules for coupling supergravity \cite{West, WB, FreedmanVanProeyen} to globally supersymmetric matter.

The next four sections briefly review what the formalism of \cite{Bergshoeff:2015tra, Dudas:2015eha, DallAgata:2015zxp, Schillo:2015ssx} implies for the couplings of supergravity to non-supersymmetric spin-zero, spin-half and spin-one particles. Although these mostly summarize known results, new material starts in \S\ref{sssec:PotStructure} and then continues in \S\ref{ssec:SMreps} with a description of how Standard Model gauge invariance can be implemented within this framework.  

\subsection{Goldstino superfield}

The central point is that the low-energy theory well below the UV supersymmetry breaking scale necessarily contains a majorana Goldstone fermion, $G$, that eventually mixes with the gravitino to give it a mass through the super-Higgs mechanism \cite{SuperHiggs}. The Goldstino provides a way for an arbitrary non-supersymmetric low-energy theory to realize supersymmetry nonlinearly. Among the points of ref.~\cite{Komargodski:2009rz} is the observation that there is no loss of generality in expressing this nonlinear realization in terms of ordinary superfields subject to constraints, with the Goldstino itself represented by a left-chiral superfield $X$ subject to a nilpotent constraint of the form $X^2 = 0$. 

In components this superfield can be schematically written in terms of the supergravity fermionic spinor coordinate $\Theta$ \cite{West, WB} by
\be \label{Xcomp}
   X = \cX + \sqrt2 \; (\Theta \gamma_\ssL G) + \Theta^2 \, F^\ssX \,,
\ee
and the nilpotent condition boils down to the constraint
\be\label{cXconstraint}
  -2 \cX \, F^\ssX + G \gamma_\ssL G = 0\,.
\ee
Provided $F^\ssX$ is a UV scale this constraint can be used to eliminate the scalar $\cX$ in terms of $G$, giving
\be
   \cX = \frac{G \gamma_\ssL G}{2F^\ssX} \,.
\ee
Using this in \pref{Xcomp} shows that this last condition is not just necessary, but also sufficient, for the condition $X^2 = 0$. 

The minimal low-energy matter sector when supersymmetry is broken consists only of the Goldstone fermion itself, for which the low-energy EFT consists of supergravity coupled to the constrained multiplet $X$. When coupled to supergravity the most general form (at the two-derivative level) for the action in superspace is \cite{DallAgata:2015zxp}
\be \label{XSugraKW}
   \cL = \int \exd^2\Theta \, 2\cE \left[ \frac38 \Bigl( \ol \cD^2 - 8 \cR \Bigr) \, e^{-K/3} + W \right] + \hbox{h.c.} \,,
\ee
where $W$ is a holormorphic function of $X$ and $K$ is a real function of $X$ and its complex conjugate $\ol X$. The most general form for these functions given the constraint $X^2 = 0$ is 
\be
   K = \ol X X   \quad \hbox{and} \quad W = \mff \,X + W_0 \,,
\ee
where the field $X$ is rescaled to choose a canonical coefficient for $\ol X X$ in $K$ and terms linear in (or independent of) $X$ in $K$ can be moved into $W$ by performing an appropriate K\"ahler transformation. 

Restoring factors of $M_p$, in the gauge $G = 0$ the component Lagrangian becomes
\be \label{XlagrComp}
   \frac{\cL}{\sqrt{-g}} = - \frac{M_p^2}2 \, \widehat R - \frac{i}2 \, \epsilon^{\mu\nu\lambda\rho} \ol \psi_\mu \gamma_5 \gamma_\nu D_\lambda \psi_\rho - \frac{1}{2M_p^2} \Bigl( W_0\, \ol \psi_\mu \gamma^{\mu\nu}\gamma_\ssL \psi_\nu + \hbox{h.c.} \Bigr) - \mff^2 + \frac{3 |W_0|^2}{M_p^2} \,,
\ee
where $\gamma^{\mu\nu} := \frac12 [ \gamma^\mu, \gamma^\nu]$ and $\widehat R = R(e,\psi)$ is the Ricci curvature, including gravitino torsion. Once the phase of $W_0$ is absorbed into the gravitino its apparent mass is revealed to be
\be\label{m32def}
   m_{3/2} = \frac{|W_0|}{M_p^2} \,,
\ee
showing the $M_p$ suppression that allows the graviton multiplet to be split by less than are other multiplets that couple more strongly to $X$.

Eq.~\pref{XlagrComp} has integrated out the auxiliary field $F^\ssX$, which given the above choices for $K$ and $W$ gives $F^\ssX = \mff$, revealing $\mff$ to be the supersymmetry breaking scale. This interpretation is also evident from the form of the scalar potential, which is
\be \label{Vsimp}
   V =   \mff^2 - \frac{3|W_0|^2}{M_p^2}  \,.
\ee
When $\mff = 0$ eqs.~\pref{m32def} and \pref{Vsimp} reproduce the supersymmetric AdS relation between curvature and gravitino mass and when specialized to $\mff^2 M_p^2 = 3|W_0|^2$ (and so $V = 0$) they give the standard flat-space relation between the supersymmetry breaking scale $\mff$ and gravitino mass.

\subsection{Non-supersymmetric fermions}

Other non-supersymmetric particles may be similarly represented in terms of constrained supermultiplets. The ones of most interest to us are those involving the SM degrees of freedom, which include scalars, spin-half fermions and spin-one gauge bosons. We start here by formulating the couplings between a standard fermion and the goldstino multiplet. 

\subsubsection{A single Majorana fermion}

For a non-supersymmetric fermion field $\psi$ we write the constrained superfield in terms of the supergravity fermionic spinor coordinate $\Theta$ by
\be \label{Ycomp}
   Y = \cY + \sqrt2 \; (\Theta \gamma_\ssL \psi) + \Theta^2 \, F^\ssY \,,
\ee
and we seek a constraint that removes its scalar part $\cY$ (but only should be able to do so once supersymmetry breaks, and so a nilpotent field $X$ is present). The constraint that does the job is $XY = 0$, which is taken to hold at the superfield level. This implies the component constraint \cite{DallAgata:2015zxp}
\be
  -4 \cY \, F^\ssX -4 \cX \, F^\ssY + 4 (G \gamma_\ssL \psi) = 0\,,
\ee
plus several consistency conditions that follow from this.

Keeping in mind that $F^\ssX$ is a UV scale -- and eliminating $\cX$ using \pref{cXconstraint} -- this constraint implies the scalar $\cY$ is given by
\be \label{cYdef}
   \cY = \frac{1}{F^\ssX} \Bigl[ -\cX \, F^\ssY + (G \gamma_\ssL \psi) \Bigr] 
   =  \frac{1}{F^\ssX} \left[ (G \gamma_\ssL \psi) -  (G \gamma_\ssL G) \, \frac{F^\ssY}{2F^\ssX} \right] \,.
\ee
Using this in \pref{Ycomp} shows that this last condition is not just necessary, but also sufficient, for the condition that $XY = 0$. This solution is the same in supergravity and global supersymmetry because the underlying constraint is chiral and algebraic. Notice also that although $Y^2$ is nonzero, expression \pref{cYdef} implies $Y^3 = 0$, since its lowest component necessarily involves at least 3 factors of the 2-component spinor $\gamma_\ssL G$ and so must vanish.  

The most general forms for the K\"ahler potential and superpotential describing the couplings of $X$ and $Y$ to supergravity consistent with the constraints $X^2 = XY = 0$ are
\be
   K = Z_{\ssX\ssX}\ol X X + Z_{\ssY\ssY}\ol Y Y + \frac14\, \hat\mfc \, Y^2 \ol Y^2  + \left( Z_{\ssX\ssY} \ol X \, Y + \frac{\hat\mfe}{2} \,  X \ol Y^2  + \frac{\hat\mfb}{2}\,  Y \ol Y^2 + \hbox{h.c.} \right) \,,
\ee
and 
\be \label{WforXY}
  W = W_0 + \hat\mff \, X + \hat\mfg \, Y + \frac12\,\hat\mfh \, Y^2 \,,
\ee
for arbitrary parameters $\hat\mfe$, $\hat\mfb$ and $\hat\mfc$ and $\hat\mff$, $\hat\mfg$ and $\hat\mfh$. We use the freedom to rescale $X \to \alpha X$ and $Y \to \beta Y + \gamma X$ to set some of these coefficients to canonical form. These are allowed redefinitions because the constraints $X^2 = XY = 0$ imply that the same constraints remain true for the new variables as well. The choices
\be
   \beta^* \beta = \frac{1}{Z_\ssY} \,, \quad \gamma = - \frac{Z_{\ssX\ssY}}{Z_\ssY} \, \alpha \,, \quad \alpha^*\alpha = \left[ Z_\ssX - \frac{\bigl| Z_{\ssX\ssY} \bigr|^2}{Z_\ssY} \right]^{-1}
\ee
do the job, and lead to 
\be
   K = \ol X X + \ol Y Y + \frac{\mfe}2 \Bigl( X \ol Y^2 + \ol X Y^2 \Bigr) + \frac{\mfb}2 \Bigl( Y \ol Y^2 + \ol Y Y^2 \Bigr) + \frac{\mfc}4 \, Y^2 \ol Y^2 \,,
\ee
and $W = W_0 + \mff \, X + \mfg \, Y + \frac12\,\mfh \, Y^2 $ with new couplings
\bea
   &&\mfe = \frac{\alpha}{Z_\ssY}\left[ \hat \mfe - \frac{Z_{\ssX\ssY}}{Z_\ssY}\, \hat \mfb\right] \,, \quad \mfb = \frac{\hat \mfb}{Z_\ssY}\,, \quad \mfc = \frac{\hat \mfc}{Z_\ssY^2} \nn\\
   &&\mff = \alpha\left[ \hat \mff - \frac{Z_{\ssX\ssY}}{Z_\ssY} \, \hat \mfg\right] \,, \quad \mfg = \frac{\hat \mfg}{\sqrt{Z_\ssY}} \,, \quad \mfh = \frac{\hat\mfh}{Z_\ssY}\,. 
\eea

With this choice the nonzero scalar parts to the superpotential derivatives are $W_\ssX = \mff$ and $W_\ssY = \mfg$ and so the physical Goldstone fermion is proportional to the linear combination $\mff \,G + \mfg \psi$. This is a pure gauge degree of freedom that can be eliminated by going to unitary gauge, corresponding to $\mff \, G + \mfg \psi = 0$. The Lagrangian is simpler to write if the gauge freedom is instead chosen to set $G = 0$, as before, in which case $\cX = \cY = 0$ and the superfields \pref{Xcomp} and \pref{Ycomp} reduce to
\be \label{XYcompG0}
    X(G = 0) =   \Theta^2 \, F^\ssX  \quad \hbox{and} \quad Y =   \sqrt2 \; (\Theta \gamma_\ssL \psi) + \Theta^2 \, F^\ssY \,.
\ee
This gauge has the minor disadvantage that it retains mixings between $\psi$ and $\psi_\mu$ which must be diagonalized within the component Lagrangian. 

In $G=0$ gauge the Lagrangian (after integrating out the auxiliary fields) is
\bea\label{fermmix}
   \frac{\cL}{\sqrt{-g}} &=& - \frac{M_p^2}2 \, \widehat R - \frac{i}2 \, \epsilon^{\mu\nu\lambda\rho} \ol \psi_\mu \gamma_5 \gamma_\nu D_\lambda \psi_\rho - \frac12 \, \ol \psi  \Dsl \psi - \mff^2  - \mfg^2 + \frac{3 |W_0|^2}{M_p^2} - \frac{\mfg}{\sqrt 2 \, M_p} \ol \psi  \gamma^\mu \psi_\mu  \nn\\
   &&\; - \frac12\left[ \left( \mfh - \mfe\mff - \mfb\mfg \right)\, \ol \psi \gamma_\ssL \psi + \frac{W_0}{M_p^2} \, \ol \psi_\mu \gamma^{\mu\nu}\gamma_\ssL \psi_\nu  + \hbox{h.c.} \right] + \hbox{(4-fermi terms)} 
\eea
which modifies \pref{Vsimp} to include the energy $\mfg^2$ associated with nonzero $F^\ssY$:
\be \label{Vsimp2}
   V =   \mff^2 + \mfg^2 - \frac{3|W_0|^2}{M_p^2}  \,.
\ee
Eq.~\pref{fermmix} also makes the mixing between $\psi$ and $\psi_\mu$ explicit (when $\mfg\neq 0$). This mixing is removed by a field redefinition of the form $\gamma_\ssL \delta \psi_\mu = A \gamma_\ssL \gamma_\mu \psi + B \gamma_\ssL D_\mu \psi$ for suitable choices of $A$ and $B$.  Once this is done the remaining physical mass term for $\psi$ is
\be \label{fermionmass}
   \cL_{\rm mass} = - \frac12  \left[ (\mfh - \mfe\mff - \mfb\mfg) + \frac{\mfg^2}{\mff^2} \left( \mfh - \mfe\mff - \mfb\mfg -  m_{3/2} \right) \right] \ol\psi \gamma_\ssL \psi + \hbox{h.c.} \,,
\ee
where (as above) $m_{3/2} = |W_0|/M_p^2$ is the gravitino mass, which for a flat vacuum satisfies $|W_0|/M_p^2 = \sqrt{  \mff^2 + \mfg^2 }/(\sqrt 3 M_p)$. The 4-fermion interactions have the form 
\bea \label{4-fermi}
  \frac{\cL_{\rm 4-fermi}}{\sqrt{-g}} &=& \left(-\frac1{8M_p^2} + \mfc - |\mfe|^2 - |\mfb|^2\right) (\ol \psi \gamma_\ssL \psi)(\ol \psi \gamma_\ssR \psi) \\
  && \qquad\qquad\ + \frac1{4M_p^2}\Bigl[ \epsilon^{\mu\nu\lambda\rho} \ol \psi_\mu \gamma_\ssL \gamma_\nu \psi_\lambda + \ol \psi_\mu \gamma_\ssL \gamma^\rho \psi^\mu \Bigr]  (\ol\psi \gamma_\ssL \gamma_\rho \psi)   \,. \nn
\eea

\subsubsection{A charged Dirac fermion}

Electrically charged Dirac fermions are particularly useful when considering Standard Model fermions. A Dirac fermion contains two left-chiral fields $Y_\pm$ whose left-handed fermions destroy particles with charge $\pm e$. We imagine both fields to satisfy the same constraint and so demand 
\be
    X^2 = X Y_+ = X Y_- = 0 \,,
\ee
and ask the Lagrangian to be invariant under the rotations $Y_\pm \to e^{\pm i \omega} Y_\pm$ with all other fields (in particular $X$) being invariant. 

In this case (after field redefinitions) the most general possible functions $K$ and $W$ describing the couplings of $Y_\pm$ to one another and to $X$  are
\be \label{WforXYchg}
  W = W_0 + \mff \, X  +  \mfm \, Y_+ Y_- \,,
\ee
and
\bea \label{KforXYchg}
   K &=& \ol X X + \ol Y_+ Y_+ + \ol Y_- Y_- + \mfe   \Bigl( \ol X Y_+ Y_-  + X \ol Y_+ \ol Y_- \Bigr) \nn\\
   && \qquad\qquad\qquad +  \mfc_{++}  \, Y_+ ^2 \ol Y_+^2 +  \mfc_{+-}  \, Y_+  Y_- \ol Y_+ \ol Y_- +  \mfc_{--}  \, Y_-^2\, \ol Y_-^2 \,.
\eea
Notice in particular that electric charge conservation prevents having a term that is linear in $Y_\pm$, and so $\mfg_\pm = 0$ and $\psi_\pm$ cannot mix with the goldstino $G$. This is because charge conservation precludes the auxiliary fields $F^{\ssY\pm}$ from acquiring expectation values.

The component Lagrangian in this case takes is 
\bea \label{DiracFL}
   \left. \frac{\cL}{\sqrt{-g}} \right|_{\mfg=0} &=& - \frac{M_p^2}2 \, \widehat R - \frac{i}2 \, \epsilon^{\mu\nu\lambda\rho} \ol \psi_\mu \gamma_5 \gamma_\nu D_\lambda \psi_\rho -  \ol \psi  \Dsl \psi - \mff^2  + \frac{3 |W_0|^2}{M_p^2}  \\
   &&\; - \left[ \left( \mfm - \mfe\mff  \right)\, \ol \psi \gamma_\ssL \psi + \frac{W_0}{2M_p^2} \, \ol \psi_\mu \gamma^{\mu\nu}\gamma_\ssL \psi_\nu  + \hbox{h.c.} \right] + \hbox{(4-fermi terms)}   \nn
\eea
where $\psi$ without a subscript is the Dirac spinor whose left- and right-handed projections are $\gamma_\ssL \psi = \gamma_\ssL\psi_-$ and $\gamma_\ssR \psi = \gamma_\ssR \psi_+$, and so 
\be
   \frac12 \Bigl(\ol \psi_+ \gamma_\ssL \psi_- + \ol \psi_- \gamma_\ssL \psi_+ + \hbox{h.c.}\Bigr)  = \ol \psi_+ \gamma_\ssL \psi_- + \ol \psi_+ \gamma_\ssR \psi_- = \ol \psi \, \psi \,.
\ee
The detailed form of the 4-fermi interactions is not needed in what follows, but it includes the terms quartic in the gravitino contained in $\widehat R$, as well as terms biquadratic in $\psi$ and $\ol\psi$ and terms involving a bilinear $\ol \psi \Gamma \psi$ (for some choice of Dirac matrices $\Gamma$) multiplying a bilinear in the gravitino. 

\subsection{Non-supersymmetric gauge boson}

The effective Lagrangian for a gauge boson is obtained in a similar way. The superfield that represents a massless spin-one particle is a left-chiral left-handed spinor, $\cW$, that contains the field strength $F_{\mu\nu} = \partial_\mu A_\nu - \partial_\nu A_\mu$. This is related to a real scalar supermultiplet, $P$, containing the gauge potential $A_\mu$ by
\be
   \cW = - \frac14 \, \Bigl( \ol\cD^2 - 8 \cR \Bigr) \cD P \,.
\ee
In terms of this the gauge kinetic term is given by the chiral integral
\be\label{GaugeKin}
   \frac{\cL}{\sqrt{-g}} = - \frac1{4g^2} \int \exd^2 \Theta \Bigl( 2 \cE \, \cW^2 + \hbox{h.c.}\Bigr) \,,
\ee
where $g$ is the gauge coupling. 

The constraint that removes the the gauge boson's fermion superpartner is given by
\be \label{XcWConstraint}
     X \cW = 0
\ee
where $X$ is (as usual) the nilpotent goldstino multiplet. Solving this constraint\footnote{What is used here is the result obtained when $X^2 = 0$ is solved first, and then $X \cW = 0$. A more general solution for which $X$ is a function of $\cW^2$ is also possible, and describes the $\cN=1$ formulation of the goldstino from a second supersymmetry should this have existed.} implies the would-be gaugino field, $\lambda$, satisfies a component constraint that allows it to be eliminated in terms of the Goldstino $G$, the gauge field $F_{\mu\nu}$, the new gauge auxiliary field D and (unlike for the previous cases) the auxiliary fields from the supergravity multiplet. The solution is complicated to write -- see \cite{DallAgata:2015zxp} -- but in the gauge $G = 0$ the spinor superfield $\cW$ becomes 
\be
   \cW = \Theta \, {\rm D} -i \gamma^{\mu\nu} \Theta \; \widehat D_\mu A_\nu + \frac{1}{M_p} \Theta^2 \, \gamma^\mu \Bigl[ \frac{i}{2} \psi_\mu \, {\rm D} - \gamma^{\lambda\rho} \psi_\mu \, \widehat D_\lambda A_\rho \Bigr]  \,,
\ee 
where the supercovariant derivative $\widehat D_\mu A_\nu$ satisfies
\be
    \widehat D_\mu A_\nu - \widehat D_\nu A_\mu  =  F_{\mu\nu} + \frac{1}{4M_p} \Bigl( \ol \psi_\mu \gamma_\nu \lambda -  \ol \psi_\nu \gamma_\mu \lambda \Bigr) \,.
\ee

The $X$-dependent generalization of the gauge kinetic term \pref{GaugeKin} then has the form
\be\label{GaugeKinX}
   \frac{\cL}{\sqrt{-g}} = - \frac1{4} \int \exd^2 \Theta \Bigl[ 2 \cE \, J (X) \, \cW^2 \Bigr] \,,
\ee
where $J(X) = J_0 + J_1 X$ is the most general holomorphic function of $X$. However the constraint \pref{XcWConstraint} implies that the $J_1 X$ term does not contribute in \pref{GaugeKinX}, and so the most general coupling of $X$ to $\cW$ has the form $\cL = \cL_\ssX + \cL_\ssW$ where $\cL_\ssX$ is given by \pref{XSugraKW} and $\cL_\ssW$ is given by \pref{GaugeKin}.  The Lagrangian in component form after elimination of auxiliary fields then is ({\it c.f.}~eq.~\pref{XlagrComp})
\bea \label{XWlagrComp}
   \frac{\cL}{\sqrt{-g}} &=& - \frac{M_p^2}2 \, \widehat R - \frac{i}2 \, \epsilon^{\mu\nu\lambda\rho} \ol \psi_\mu \gamma_5 \gamma_\nu D_\lambda \psi_\rho - \frac{1}{4 e_g^2} \, F_{\mu\nu} F^{\mu\nu}  \\
   && \qquad\qquad - \frac{1}{2M_p^2} \Bigl( W_0\, \ol \psi_\mu \gamma^{\mu\nu}\gamma_\ssL \psi_\nu + \hbox{h.c.} \Bigr) - \mff^2 + \frac{3 |W_0|^2}{M_p^2} \,,\nn
\eea
which uses $J_0 = 1/e_g^2$.

\subsection{Non-supersymmetric scalar}

The final practical example is the case of non-supersymmetric scalar fields that have no fermionic superpartners. Following \cite{Komargodski:2009rz} we consider first a complex scalar and restrict to its real and imaginary components at the end.

\subsubsection{Complex scalar}

A complex scalar appears as the lowest component of a chiral superfield  
\be
   H = \cH + \sqrt2 \, \Theta \gamma_\ssL \Psi + \Theta^2 \, F^\ssH \,,
\ee
and the constraint that implements the nonlinear realization is the condition that the product $X \ol H$ be left-chiral, and so
\be \label{ScalarConst}
    \ol\cD\Bigl( X \ol H \Bigr) = 0 \,,
\ee
where $\cD$ is the left-handed superspace spinor covariant derivative. This constraint turns out to allow both $\Psi$ and $F^\ssH$ to be eliminated in terms of other fields, leaving only $\cH$ as a physical propagating degree of freedom. 

For the purposes of writing down invariant Lagrangians it is noteworthy that \pref{ScalarConst} also implies $\ol\cD(X \,\ol H^n) = 0$ for any power $n$ and therefore
\be\label{FXconstraint}
    \ol\cD \Bigl[ X \, \cF(H, \ol H) \Bigr] = 0
\ee
for any function $\cF(H,\ol H)$ of $H$ and its conjugate. This states that the constraints ensure that arbitrary functions of $H$ and $\ol H$ become left-chiral once multiplied by $X$. 

The constraints $X^2 = 0$ and $\ol\cD(X \ol H) = 0$ can be used to eliminate $\Psi$ and $F^\ssH$ from $H$, leading to complicated expressions that simplify considerably in the gauge $G = 0$, for which
\be \label{XHinG0}
    X = \Theta^2 \, F^\ssX \quad \hbox{and} \quad H = \cH \qquad \hbox{(if $G = 0$)}\,.\
\ee
The absence of $F^\ssH$ in \pref{XHinG0} means that there is no need to integrate out $F^\ssH$ when constructing the component Lagrangian, and so the result (given below) is {\it not} simply the standard 4D sugra Lagrangian for a chiral field $H$ with the fermionic partner for $\cH$ set to zero.  

As usual, the two-derivative Lagrangian is specified by the functions $K(X,H,\ol X , \ol H)$ and $W(X,H,\ol H)$, and in the present case the most general form consistent with the constraints is
\be \label{KgenScalar}
   K = \cU(H,\ol H) \, X \ol X + X \hat \mfP(H,\ol H) + \ol X \, \hat{\ol \mfP}(H , \ol H) + \mfZ(H,\ol H) 
\ee
and
\be \label{WgenScalar}
   W = \mfw_0(H) + X \,\hat{\mfw}_{\ssX}(H,\ol H) \,.
\ee
Also as usual, there is considerable freedom to simplify this form using field redefinitions. For instance, the function $\cU(H,\ol H)$ premultiplying $X\ol X$ can be rescaled into a redefinition $X \to \widehat X = \sqrt{\cU} \; X$. This is possible even if $\cU$ depends on $\ol H$ because the constraint \pref{FXconstraint} implies $\widehat X$ remains chiral, and the nilpotent condition $X^2 = 0$ implies the same remains true for the new variable: $\widehat X^2 = 0$. This means that any nonzero $\cU$ can be absorbed into the combinations $\mfP = \hat\mfP/\sqrt{\cU}$ and $\mfw_\ssX = \hat\mfw_\ssX/\sqrt{\cU}$. Performing a K\"ahler transformation similarly shows that $\mfP$ appears through the combination $ \mfw_{\ssX} + ({\mfP \, \mfw_0}/{M_p^2})$.  

The component Lagrangian (in the gauge $G=0$) then becomes
\bea \label{XHlagrComp}
   \frac{\cL}{\sqrt{-g}} &=& - \frac{M_p^2}2 \, \widehat R - \frac{i}2 \, \epsilon^{\mu\nu\lambda\rho} \ol \psi_\mu \gamma_5 \gamma_\nu D_\lambda \psi_\rho - \mfZ_{\ssH\ol\ssH} \, \partial_\mu \ol \cH \,\partial^\mu \cH - V(\cH,\ol \cH) \\
   && \qquad\qquad + \left[\frac1{4M_p^2} \, \epsilon^{\mu\nu\lambda\rho}   \ol \psi_\mu \gamma_\ssR \gamma_\nu \psi_\lambda \, \mfZ_\ssH\,  \partial_\rho \cH - \frac{1}{2} \, \mfM(\cH,\ol\cH)\, \ol\psi_\mu \gamma^{\mu\nu} \gamma_\ssL \psi_\nu + \hbox{h.c.} \right] \,,\nn
\eea
where $\mfZ_{\ssH\ol\ssH} = \partial_\ssH \partial_{\ol\ssH}\mfZ$ controls the scalar kinetic function. The functions $\mfM(\cH,\ol \cH)$ and $V(\cH, \ol\cH)$ are defined by
\be \label{Mfunction}
    \mfM  := \frac{\mfw_0}{M_p^2} \, e^{\mfZ/(2M_p^2)}   \,,
\ee
and
\be \label{FtermNilpotent0}
   V(\cH,\ol \cH) := e^{\mfZ/M_p^2} \left[ \left|  \mfw_{\ssX} + \frac{\mfP \, \mfw_0}{M_p^2} \right|^2 - \frac{3|\mfw_0|^2}{M_p^2} \right] = e^{\mfZ/M_p^2} \left[ \frac{1}{\cU}\left| \hat\mfw_{\ssX} + \frac{\hat\mfP \, \mfw_0}{M_p^2} \right|^2 - \frac{3|\mfw_0|^2}{M_p^2} \right]\,.
\ee
In detail, the factors of $e^{\mfZ/M_p^2}$ come from the Weyl rescalings needed to put the Einstein term into canonical form, and the remaining potential arises as a combination of squares due to the elimination of the supergravity auxiliary fields as well as $F^\ssX$.

\subsubsection{Real scalar}

We remark in passing that ordinary real scalar fields are represented by superfields $B$ satisfying the constraint
\be \label{RealOrdinaryScalar}
   X \Bigl( B - \ol B \Bigr) = 0
\ee
since this allows the imaginary part of the scalar $\cB \in B$ to be eliminated in terms of the Goldstone fermion and other fields. Notice that \pref{RealOrdinaryScalar} also implies the constraint \pref{ScalarConst} given above. Solving \pref{RealOrdinaryScalar} also shows that it implies a subsidiary condition $(B - \ol B)^3 = 0$, which can be helpful when exploring how $B$ can appear in the action \cite{Komargodski:2009rz, Bergshoeff:2015tra, Dudas:2015eha, DallAgata:2015zxp, Schillo:2015ssx, Inflation}.

\subsubsection{Structure of the scalar potential}
\label{sssec:PotStructure}

Eq.~\pref{FtermNilpotent0} is surprising because it preserves the remarkably specific supergravity-type scalar-potential form despite supersymmetry being so badly broken that the scalar's fermionic partner can be integrated out. In particular it is strictly positive in the limit $M_p \to \infty$. But we are normally free to write down arbitrary potentials $U(\cH,\ol\cH)$ for non-supersymmetric scalars, so how is this consistent with the scalar potential taking the form \pref{FtermNilpotent0}? 

When answering this question it is instructive first to consider what happens in the limit of global symmetry. In global supersymmetry the Lagrangian is obtained by integrating $K$ and $W$ from \pref{KgenScalar} and \pref{WgenScalar} over the fermionic coordinates
\be
  \cL = \int \exd^2\theta \, \exd^2 \ol \theta \; K + \left[ \int \exd^2 \theta \; W + \hbox{h.c.} \right] \,,
\ee
and in this language it is the contributions $\cU(H,\ol H) \ol X X \subset K$ and $\hat \mfw_\ssX(H,\ol H) X \subset W$ that give a generic potential\footnote{Because $X \hat \mfP(H,\ol H)$ is left-chiral it gives a total derivative once integrated over $\exd^2 \theta \, \exd^2 \ol\theta$, and so in global supersymmetry drops out of the scalar potential completely.}
\be \label{VvsFx}
  V =- \left| F^\ssX \right|^2 \cU(\cH,\ol\cH) +\Bigl[F^\ssX  \hat \mfw_\ssX(\cH, \ol\cH) + \hbox{h.c.} \Bigr]
\ee
and this can indeed reproduce an arbitrary potential $V(\cH,\ol\cH)$ once the auxiliary field for $X$ is replaced by a spurionic expectation value $F^\ssX = \mu^2$. 

But having supersymmetry be realized on the low-energy fields implies that $F^\ssX$ is {\it not} simply a constant spurion. It is a field over which a functional integral is performed, and once this is done for the gaussian function \pref{VvsFx} one obtains
\be
  V = \frac{\left|\hat \mfw_\ssX(\cH,\ol\cH)\right|^2}{\cU(\cH,\ol\cH)} = \bigl| \mfw_\ssX \bigr|^2
\ee
in agreement with the $M_p \to \infty$ limit of \pref{FtermNilpotent0}. The last equality rescales $X \to \widehat X = \sqrt{\cU} \; X$ -- as described below \pref{WgenScalar} -- but also underlines that this redefinition breaks down in regions where $\cU$ changes sign. The spurion limit applies best in situations where $F^\ssX$ is dominated by a large $\cH$-independent supersymmetry breaking contribution, such as when $\mfw_\ssX(\cH,\ol \cH) = \mu^2 + \mfv(\cH,\ol \cH)$ with $\mu^2 \gg |\mfv(\cH ,\ol\cH)|$, since in this case 
\be \label{LEVspurion}
  V \simeq \mu^4 + \mu^2\Bigl[ \mfv(\cH,\ol\cH) + \hbox{h.c.} \Bigr] + \bigl| \mfv(\cH,\ol\cH) \bigr|^2 \,.
\ee 
Because constant contributions to $V$ have no significance in the absence of gravity this shows in global supersymmetry how arbitrary potentials $\mfv(\cH,\ol\cH)$ with general signs can emerge for scalars when global supersymmetry is badly broken.  

Put another way: the sum/difference-of-squares structure of the potential \pref{FtermNilpotent0} arises because of the presence of auxiliary fields in the low-energy theory; both $F^\ssX$ and the auxiliary fields of the gravity multiplet itself. Although these fields do not propagate they are also not optional if supersymmetry is to be realized linearly by the constrained field content, given the assumption that the goldstino, the graviton and gravitino are light enough to be in the low-energy effective theory. Conversely, if one were to try to couple supergravity to a potential not of the sum/difference-of-squares form the absence of auxiliary fields should prevent linearly realizing supersymmetry within the gravity sector and so mass splittings within the gravity mulltiplet should not remain small. 

This entire discussion underlines the importance of including non-propagating fields (like auxiliary fields or topological fields), particularly for naturalness arguments that depend on the form of the scalar potential in the low-energy theory. 

\subsection{Standard Model representations}
\label{ssec:SMreps}

It can be useful to identify how the Standard Model itself is described using the above fields. From the superfield point of view the Standard Model field-content assigns a constrained superfield for each known particle. It therefore consists of constrained spin-one multiplets for all of the gauge fields; a constrained fermion multiplet for every left-handed Standard Model fermion -- two electroweak doublets $L$ and $Q$ plus the electroweak singlets $E$, $U$ and $D$; and a constrained scalar multiplet $H$ containing the Higgs doublet. 

Comparing with the fermion mass terms of \pref{DiracFL} shows that the `down-type' fermion Yukawa couplings appear within the superpotential, such as for 
\be \label{SMYukInW}
  \mfw_0 =  y_\ssE L E H + y_\ssD Q D H \,.
\ee
The inability to write a similar $y_\ssU Q U \ol H$ term for `up-type' fermions (because supersymmetry forbids $\ol H$ from appearing in $W$) is one of the reasons a second Higgs is needed in supersymmetric versions of the Standard Model. One might hope to use the constraint \pref{ScalarConst} to evade this problem, but that requires at least one factor of $X$ in addition to $\ol H$. 

The up-type Yukawa instead arise as part of the K\"ahler potential, through terms like
\be
    K \ni y_\ssU Q U \ol H \left( \frac{\ol X}{\mff} \right) + \hbox{h.c.}
\ee
which contributes to fermion masses by an amount $y_\ssU v$ when $\mff$ is the parameter appearing in $\mff \, X \in W$ that dominates in $F^\ssX$ ({\it c.f.}~the contribution $\mfe \mff$ in the mass term shown in \pref{fermionmass}). Because the $F$ auxiliary fields for Standard Model fermions transform under gauge transformations, they in particular must vanish in any vacuum that does not break electromagnetic $U(1)$ invariance. As a consequence Standard Model fermions do not mix with the goldstino that lives within the superfield $X$. 

Keeping in mind the discussion surrounding \pref{LEVspurion} the choice that reproduces the Higgs potential of the Standard Model when used in \pref{FtermNilpotent0} (and Planck-suppressed terms are dropped) is similarly 
\be \label{SMHiggsInW}
   \mfw_\ssX = \mff_0 + c_0 + c_1 \, \ol H H+ c_2 (\ol H H)^2 + \cdots = \mff + \frac{\lambda}{\mff} \, (\ol H H - v^2)^2 + \cdots\,,
\ee
for constants $\lambda$ and $v$, where (again) $\mff = \mu^2$ sets the supersymmetry-breaking scale. Regarding this as an expansion in powers of $1/\mff$ seems reasonable because the absence of Standard Model superpartners presumably requires $\mff \gg v^2$ since these partners have masses controlled by $\mff$ and are assumed to be much heavier than the electroweak scale.\footnote{Strictly speaking only $F^\ssX$ must be large compared to the weak scale, so assuming $\mff \gg v^2$ makes the additional assumption that $F^\ssX$ is dominated by the globally supersymmetric part $W_\ssX$ in $F^\ssX \propto W_\ssX + (K_\ssX W/M_p)$. Ref.~\cite{CCNo-Scale} explores the more counter-intuitive regime where this assumption fails, such as when $\ol H H$ seeks a vev of order $\mff_0 + c_0$ that makes $w_\ssX$ vanish in \pref{SMHiggsInW}.}

The structures described above should be generic when the gravity sector is much more supersymmetric than is the Standard Model sector, with the $X$ field being a remnant from much higher energies containing the supersymmetry-breaking order parameters. The main assumption is that only the one field $X$ required to contain the Goldstone fermion is light enough to descend from higher energies. As argued in \cite{Komargodski:2009rz} supersymmetric current algebra ensures that the $X$ field parametrizes the low-energy limit of {\it any} such supersymmetry breaking  in the UV (even, for example, if $D$-terms carry part of the breaking in the high-energy theory).  

\section{Wilsonian Flow}
\label{ssec:Naturalness}

The generality of the nonlinear realization described in \S\ref{sec:SuperNonlin} guarantees that it should describe the low energies well below the supersymmetry breaking scale, assuming only the hierarchy $\Delta M_\SM \gg \Lambda \gg \Delta M_\SG$ between the EFT's UV scale $\Lambda$ and the mass splittings $\Delta M_\SM$ and $\Delta M_\SG$ between particles and their superpartners in the Standard Model and gravity supermultiplets. But the low-energy theory nonetheless has counter-intuitive properties, such as the peculiar supergravity form of scalar potential given in \pref{FtermNilpotent0}. Why should this survive loop corrections once successive waves of non-supersymmetric heavy fields are integrated out?

It is instructive therefore to be explicit, and integrate out heavy degrees of freedom to see how the resulting threshold corrections preserve these counter-intuitive properties into the low-energy limit. The goal is to see how the defining functions like $K$ and $W$ change as energies are lowered below the fermion threshold. 

In this section our tools for doing this are standard calculations using the Gilkey-DeWitt heat-kernel coefficients \cite{DeWitt:1964mxt, HeatKernel, HKReview, HKXD}, that give the effects of integrating out heavy particles at one loop order. These tools show that such loops contribute local shifts to the effective Lagrangian of the form $\cL_\ssI = \cL_u + \delta \cL$ with
\be \label{footnoteeq}
  \frac{\delta \cL}{\sqrt{-g}} = a_{cc}^{(s)} m^4 +  a_{eh}^{(s)} \, m^2 \, R + \frac{a_{rs}^{(s)} m^2}{M_p^2}\, i\epsilon^{\mu\nu\lambda\rho} \ol \psi_\mu \gamma_5 \gamma_\nu D_\lambda \psi_\rho + \frac{a_{gm}^{(s)} m^3}{M_p^2} \, \ol \psi_\mu \gamma^{\mu\nu} \psi_\nu +  \cdots
\ee
where $a_{cc}^{(s)}$, $a_{eh}^{(s)}$, $a_{rs}^{(s)}$ and $a_{gm}^{(s)}$ are dimensionless order $1/(16\pi^2)$ quantities whose values -- listed explicitly in \cite{HeatKernel, HKReview, HKXD} -- depend on the spin $s$ of the particle that was integrated out. They can also depend logarithmically\footnote{The logarithmic mass-dependence that enters through loop corrections plays an important role in \cite{CCNo-Scale}, since it can introduce a logarithmic dependence on some of the low-energy fields.} on the heavy-particle mass $m$. The $a_{rs}^{(s)}$ and $a_{gm}^{(s)}$ terms arise in a Planck-suppressed way because each gravitino vertex comes with a power of $1/M_p$ (see for example the 2-fermi interactions in \pref{fermmix} or \pref{DiracFL} or the 4-fermi interactions of \pref{4-fermi}). The factors of $m$ in \pref{footnoteeq} evaluate the loop using dimensional regularization renormalized using minimal subtraction so that the mass of the particle in the loop provides the largest mass scale within the loop integral and so set its dimensions. 

\subsection{Integrating out a massive Dirac fermion}
\label{sssec:IntOutMassDirac}

Consider first integrating out a single massive Dirac fermion. Broadly speaking there are two situations for which massive non-supersymmetric fermions might arise, depending on whether or not the fermion's bosonic superpartner is lighter or heavier than is the fermion itself. A common case has the bosonic partner heavier than the fermion, such as when the boson is a scalar with no symmetries that protect its mass, so we consider this case first. 

\subsubsection{Only supergravity at low energies}

If the boson were heavier it would be integrated out first when coming down in scales, leaving a low-energy EFT involving only the fermion and the supergravity sector. In this section we see how integrating out the remaining fermion is captured by threshold corrections to the functions $W$ and $K$ governing the low-energy nonlinear realization. 

Since the fermion is a charged Dirac particle the effective Lagrangian that applies above its mass is given at the two-derivative level by the superpotential $W$ and K\"ahler potential $K$ given in \pref{WforXYchg} and \pref{KforXYchg}, repeated here for convenience:
\be \label{KWforXYchg1}
  W_\UV = W_0 + \mff \, X  +  \mfm \, Y_+ Y_- \,,
\ee
and
\bea \label{KWforXYchg2}
   K_\UV &=& \ol X X + \ol Y_+ Y_+ + \ol Y_- Y_- + \mfe   \Bigl( \ol X Y_+ Y_-  + X \ol Y_+ \ol Y_- \Bigr) \\
   && \qquad\qquad\qquad\qquad\qquad +  \mfc_{++}  \, Y_+ ^2 \ol Y_+^2 +  \mfc_{+-}  \, Y_+  Y_- \ol Y_+ \ol Y_- +  \mfc_{--}  \, Y_-^2 \ol Y_-^2 \,.\nn
\eea
The corresponding component Lagrangian \pref{DiracFL}, also repeated here for convenience, is:
\bea \label{DiracFLx}
     \frac{\cL_u}{\sqrt{-g}}   &=& - \frac{M_p^2}2 \, \widehat R - \frac{i}2 \, \epsilon^{\mu\nu\lambda\rho} \ol \psi_\mu \gamma_5 \gamma_\nu D_\lambda \psi_\rho -  \ol \psi  \Dsl \psi -  \mff^2  + \frac{3 | W_{0}|^2}{M_p^2}  \\
   &&\; - \left[ \left( \mfm - \mfe \mff  \right)\, \ol \psi \gamma_\ssL \psi + \frac{W_{0}}{2M_p^2} \, \ol \psi_\mu \gamma^{\mu\nu}\gamma_\ssL \psi_\nu  + \hbox{h.c.} \right] + \hbox{(4-fermi terms)} \,. \nn
\eea
The terms quadratic in fields are characterized by parameters $\mff$, $W_{0}$ and $m = \mfm - \mfe \, \mff$.
 
Below the fermion mass the corresponding terms of the effective Lagrangian are instead described by \pref{XlagrComp} and so are characterized by $K_\IR = \ol X X$ and the IR parameters $\tilde\mff$ and $\widetilde W_{0}$ appearing in $W_\IR = \widetilde W_0 + \tilde \mff X$. In components this gives
 \be \label{XlagrCompx}
   \frac{\cL_\ssI}{\sqrt{-g}} = - \frac{M_p^2}2 \, \widehat R - \frac{i}2 \, \epsilon^{\mu\nu\lambda\rho} \ol \psi_\mu \gamma_5 \gamma_\nu D_\lambda \psi_\rho - \frac{1}{2M_p^2} \Bigl( \widetilde W_0\, \ol \psi_\mu \gamma^{\mu\nu}\gamma_\ssL \psi_\nu + \hbox{h.c.} \Bigr) - \tilde \mff^2 + \frac{3 |\widetilde W_0|^2}{M_p^2} \,.
\ee
The goal is to compute what matching at the fermion threshold implies for how the parameters $\widetilde W_0$ and $\tilde \mff$ depend on their UV counterparts $W_0$, $\mff$ and $m$.

Integrating out a heavy Dirac fermion shifts the effective Lagrangian with the local correction given as a special case of \pref{footnoteeq}, with coefficients $a^{(1/2)}_{cc}$, $a^{(1/2)}_{eh}$, $a^{(1/2)}_{rs}$ and $a^{(1/2)}_{gm}$ specialized to spin $s = 1/2$. For brevity of notation we suppress the superscript $(s)$ in what follows.

The contributions involving $a_{eh}$ and $a_{rs}$ change the canonical normalization of the metric and gravitino fields and so are absorbed into field redefinitions
\be
   g_{\mu\nu} \to \lambda_g \, g_{\mu\nu} \quad \hbox{and} \quad
   \psi_\mu \to \lambda_f \, \psi_\mu \,.
\ee
Preserving the form of the Einstein-Hilbert part of the action implies
\be
   \lambda_g = \left( 1 - \frac{2\, a_{eh} m^2}{M_p^2} \right)^{-1} \simeq 1 + \frac{2\, a_{eh} m^2}{M_p^2} + \cdots\,,
\ee
and this metric redefinition then rescales all of the other terms in the Lagrangian because $\sqrt{-g} \to \lambda_g^2 \sqrt{-g}$, $\epsilon^{\mu\nu\lambda\rho} \to \lambda_g^{-2} \epsilon^{\mu\nu\lambda\rho}$ and $\gamma_\mu \to \lambda_g^{1/2} \gamma_\mu$ {\it etc}. Preserving the form of the gravitino kinetic term therefore requires
\be
   \lambda_f = \lambda_g^{-1/4} \left( 1 - \frac{2a_{rs} m^2}{M_p^2} \right)^{-1/2} \simeq 1 + \frac{m^2}{M_p^2}\left( a_{rs} - \frac{a_{eh}}{2} \right) + \cdots  \,.
\ee

With these conventions the coefficient of the gravitino mass term in $\cL_u + \delta \cL$ becomes
\bea
  \cL_u + \delta \cL &\ni& - \frac{\lambda_f^2 \lambda_g}{2M_p^2} \sqrt{-g}\, (\ol \psi_\mu \gamma^{\mu\nu} \psi_\nu )\left(W_0  - 2a_{gm} m^3  \right) \nn\\
  &\simeq& - \frac{1}{2M_p^2} \sqrt{-g}\, (\ol \psi_\mu \gamma^{\mu\nu} \psi_\nu )\left(W_0 - 2a_{gm} m^3  \right) \left[1 + \frac{m^2}{M_p^2} (a_{eh} + 2 a_{rs}) +\cdots \right]
\eea
so comparing this to \pref{XlagrCompx} allows us to identify the value for the IR parameter $\widetilde W_0$:
\be \label{W0shift}
    \widetilde W_0 \simeq \left(W_0 - 2a_{gm} m^3  \right) \left[1 + \frac{m^2}{M_p^2} (a_{eh} + 2 a_{rs}) +\cdots \right] \simeq  W_0 - 2a_{gm} m^3 + \hbox{($M_p$ suppressed)} \,.
\ee

Repeating this exercise for the vacuum energy $\cL_u + \delta \cL \ni -\sqrt{-g}\; V$ gives
\be
  V \simeq \lambda_g^2  \left(\mff^2 - \frac{3|W_0|^2}{M_p^2}  - a_{cc} m^4 \right) \simeq  \left( \mff^2 - \frac{3| W_0|^2}{M_p^2}   - a_{cc} m^4 \right)\left( 1 + \frac{4\, a_{eh} m^2}{M_p^2} + \cdots \right) \,,
\ee
and comparing this to $\mff^2 - 3|W_0|^2/M_p^2$ fixes the value for the IR parameter $\tilde \mff$:
\bea
  \tilde \mff^2 &=& V + \frac{3|\widetilde W_0|^2}{M_p^2} \nn\\
  &\simeq& \left( \mff^2 - \frac{3| W_0|^2}{M_p^2}   - a_{cc} m^4 \right)\left( 1 + \frac{4\, a_{eh} m^2}{M_p^2} + \cdots \right) \\
  &&\qquad\qquad\qquad + \frac{3}{M_p^2}\left| W_0 - 2a_{gm} m^3  \right|^2 \left[1 + \frac{2m^2}{M_p^2} (a_{eh} + 2 a_{rs}) + \cdots\right] \nn\\
  &\simeq& \mff^2 - a_{cc} m^4 + \hbox{($M_p$-suppressed)} \,.\nn
\eea
Consequently (assuming $\mff^2 \gg a_{cc} m^4$) the IR theory's superpotential becomes\footnote{Having the superpotential $W$ be renormalized when the heavy field is integrated out apparently contradicts the supersymmetric non-renormalization theorems. This is possible when supersymmetry is nonlinearly realized because the lack of kinetic terms for the Lagrange multipliers means their propagators do not take the form assumed when these theorems are proven.} 
\be \label{mffshift}
 W_\IR = \widetilde W_0 + \tilde \mff \, X \simeq \Bigl( W_0 - 2a_{gm} m^3 \Bigr) + \left(\mff  - \frac{a_{cc} m^4}{2 \mff} \right) X + \cdots \,.
\ee
Comparing to \pref{XlagrCompx} assumes that any changes to $K$ are absorbed into rescalings of $X$ so that $K_\IR = \ol X X$. In all of these expressions the physical fermion mass is related to the UV parameters by $m^2 = |\mfm - \mfe \mff|^2$, as described above.

\subsubsection{Residual light scalar}

Consider next the case where the low-energy theory below the fermion mass contains a non-supersymmetric scalar in addition to the supergravity sector. In this case the low energy theory requires more than just the parameters $\widetilde W_0$ and $\tilde\mff$, being replace by expressions like \pref{XHlagrComp}, repeated here as
\bea \label{XHlagrCompx}
   \frac{\cL_\ssI}{\sqrt{-g}} &=& - \frac{M_p^2}2 \, \widehat R - \frac{i}2 \, \epsilon^{\mu\nu\lambda\rho} \ol \psi_\mu \gamma_5 \gamma_\nu D_\lambda \psi_\rho - \tilde\mfZ_{\ssH\ol\ssH} \, \partial_\mu \ol \cH \,\partial^\mu \cH - \widetilde V(\cH,\ol \cH) \\
   && \qquad\qquad + \left[\frac1{4M_p^2} \, \epsilon^{\mu\nu\lambda\rho}   \ol \psi_\mu \gamma_\ssR \gamma_\nu \psi_\lambda \, \tilde \mfZ_\ssH\,  \partial_\rho \cH - \frac{1}{2} \,\widetilde \mfM(\cH,\ol\cH)\, \ol\psi_\mu \gamma^{\mu\nu} \gamma_\ssL \psi_\nu + \hbox{h.c.} \right] \,,\nn
\eea
and something similar happens for the UV lagrangian, whose parameters $\mff$ and $W_0$ are also replaced by $\cH$-dependent quantities $\mfw_\ssX(\cH, \ol \cH)$ and $\mfw_0(\cH)$.

The lagrangian shift $\delta \cL$ due to integrating out the fermion field is again given by an expression like \pref{footnoteeq}. When matching between UV and IR theories the comparison of parameters is simplest when done at order $M_p^0$, leading to 
\be \label{mffshift2}
 W_\IR = \tilde\mfw_0 + \tilde\mfw_\ssX \, X \simeq \Bigl( \mfw_0 - 2a_{gm} m^3 \Bigr) + \left(\mfw_\ssX  - \frac{a_{cc} m^4}{2 \mff} \right) X + \cdots \,,
\ee
where $m$ is now $\cH$-dependent. Comparing this to expressions like \pref{SMHiggsInW} -- and keeping in mind the logarithms of $m$ that are implicit in quantities like $a_{cc}$ -- shows how the low-energy scalar potential acquires its standard $m^4 \ln (m^2/\mu^2)$ Coleman-Weinberg type \cite{ColemanWeinberg} corrections, as might have been expected.

\subsection{Other spins}

The generality of the result \pref{footnoteeq} shows that similar considerations apply when integrating out other fields with different spins. In more general cases the shifts to the superpotential parameters like $\mff$ and $W_0$ involve a sum over $s$. When integrating out multiple fields one finds similar expressions as above for these parameters, but summed over the masses and couplings of all of the heavy states that are integrated out. 

The sum over $s$ can introduce cancellations between contributions from particles with different spins, such as when the integrated-out particles happen to flesh out a complete supermultiplet whose average mass is larger than the mass splittings amongst its members. The sum over elements of the multiplet then tend to give contributions proportional to spin-weighted supertraces of the mass matrix, with (for instance) the $m^4$ contributions to \pref{mffshift2} combining (when summed over spins) into STr $M^4 = \sum_s (-)^{2s}(2s+1) m^4_s$. Known mass sum rules for spontaneously broken supersymmetry (see for instance \cite{SuperTrace, SuperHiggs} and \cite{Ferrara:1994kg, DeAlwis:2021gou}), then allow these to cancel, as is required by the non-renormalization theorems \cite{NRTheorems}. 

Auxiliary fields ({\it e.g.} for $X$) enter into these calculations through the supersymmetry breaking mass splittings within a heavy multiplet. In the supersymmetric limit they play little or no role, but they dominate for badly split multiplets. This is in detail why they drop out of supersymmetric radiative corrections while non-supersymmetric heavy particles dominantly contribute to the $X$-dependent part of the action. It is ultimately the role of the auxiliary fields to allow the generic component form be expressible in terms of the specific kinds of couplings ({\it i.e.}~$W$ and $K$) that appear in the superspace Lagrangian.

\section{Concluding Remarks}
\label{sec:Conclusions}

We conclude by summarizing the main arguments, and briefly discussing directions towards which this point of view likely leads.

\subsection{Summary of main results}
Our main point is to remove a relatively widespread objection to the use of supergravity in very low-energy applications, such as to astrophysics and post-nucleosynthesis cosmology. Although supergravity has been used to construct many such models, particularly for Dark Energy in cosmology, two related objections have been raised that may have prevented its wider exploration:
\begin{itemize}
\item The ingredients of interest in applications -- usually small scalar masses and small vacuum energies -- are controlled by the low-energy scalar potential and this is known to be particularly sensitive to quantum corrections involving the theory's heavy particles;
\item None of the known heavy particles relevant to cosmology or astrophysics are supersymmetric and so their quantum effects are likely to badly break any supersymmetric structure even if this were present at higher energies.
\end{itemize}

Recent developments describing broken supersymmetries in terms of constrained superfields \cite{Komargodski:2009rz, Bergshoeff:2015tra, Dudas:2015eha, DallAgata:2015zxp, Schillo:2015ssx} show how supergravity couples to a matter sector in which supersymmetry is badly broken, and so allow the explicit calculation of the non-supersymmetric quantum effects required to test this objection more precisely. We use this formalism to provide an estimate of their size and argue in favour of the stability against loop corrections of the structure predicted by \cite{Komargodski:2009rz, Bergshoeff:2015tra, Dudas:2015eha, DallAgata:2015zxp, Schillo:2015ssx}.

This opens up a conceptual framework that is attractive in its own right and so deserves more systematic study: the vision that high-energy supersymmetry survives at low energies dominantly in the (possibly complicated) gravity sector despite supersymmetry being badly broken for all of the ordinary particles described by the Standard Model. This is a vision that often actually does descend from supersymmetric UV completions \cite{SLED, HighEUV} (though the calculations that show this are usually only done without including quantum corrections). It does so because splittings in any supermultiplet arise proportional to the couplings of that multiplet to the supersymmetry-breaking order parameter, and gravity usually has the weakest couplings of all. 

Besides stability against loops, there embedding of the Standard Model into supersymmetry described here differs in several potentially useful ways relative to the standard MSSM approach.\footnote{Note that our scenario differs from the non-linear MSSM -- proposed in \cite{Antoniadis:2010hs} with the idea of extracting model independent couplings of the goldstino to the MSSM -- for which the nilpotent superfield couples to the standard MSSM field content (for which the Standard Model superfields realize supersymmetry linearly). See also \cite{Li:2020rzo} for a proposal closer to ours.}
\begin{enumerate}
\item The functional dependence of the scalar potential keeps the its supergravity structure as a difference of perfect squares (such as in eq.~\pref{FtermNilpotent0}). We argue that this property follows from the low-energy presence of the auxiliary fields for the nilpotent superfield $X$  and for the supergravity sector itself (for which supersymmetry is linearly realized).

\item At the low energies of interest here all Standard Model partners have been integrated out and so (for example) pose no problem with anomalies in the low-energy EFT (removing one of the MSSM arguments that one needs a second Higgs superfield). For down quarks and leptons Yukawa couplings arise in the usual MSSM way within the superpotential $W$, but the Yukawa couplings for the up quarks come from their coupling to the $X$ superfield in the K\"ahler potential. It is the presence of the nilpotent superfield $X$ that allows one to evade MSSM arguments and have Yukawa couplings for both up and down quarks with only a single Higgs. 

\item The Higgs potential also arises from the coupling of the Higgs $H$ to the nilpotent superfield $X$ within the superpotential. Because $XH\ol H$ is dimension 3, it can only have dimensionless coefficients and there is no $\mu$ problem. It is the constraint satisfied by the Higgs superfield that ensures the quantity $X\cH\ol\cH$ is chiral and so allows $\ol\cH$ to appear in the superpotential.

\end{enumerate}

\subsection{Phenomenological speculations} 

A detailed phenomenological study of this scenario is beyond the scope of this article, but we close with some speculations about how this picture of nature -- {\it i.e.}~a very supersymmetric gravity sector coupled to a non-supersymmetric matter sector -- might impact some of the puzzling questions of our time.

\subsubsection*{Planck-coupled Dark Sector}

The most obvious consequence of this framework is the inevitable complication of the gravity sector, which at the very least must contain a gravitino. Explicit examples often involve other gravitationally coupled supermultiplets, such as by including a low-energy dilaton-axion multiplet arising from the accidental scale invariances that are ubiquitous to higher-dimensional supergravity \cite{Salam:1989fm, Burgess:2011rv, SUGRAscaleinv, BMvNNQ, GJZ} (and string theory \cite{Burgess:2020qsc}).

This suggests the existence of a rich spectrum of gravitationally coupled particles, although one whose properties are constrained by supersymmetry. Such a sector comes with potentially observational implications (and constraints) coming from cosmology and astrophysics. The best known of these are the cosmological bounds on the presence of a light gravitino or other superpartner (or moduli, in explicit UV completions) \cite{Kawasaki:2008qe, PDG, Feng:2010gw, Coughlan:1983ci, Banks:1993en, deCarlos:1993wie, Conlon:2007gk}. Although these constrain allowed gravitino properties they are also not excluded over a wide mass range.

Having a potentially large number of dark-sector particles also underlines the importance of studying the three renormalizable `portals' --- Higgs-scalar, gauge-kinetic and neutrino --- through which Standard Model singlets (like Dark Matter) can interact with Standard Model fields without suppression by heavy mass scales (for Dark Matter models that exploit the scalar portal in this way see {\it e.g.}~\cite{ScalarPortal}). 

\subsubsection*{Light sterile fermions}

A special case of the general observation that the gravity sector might be complicated is the supersymmetric requirement that it must contain gravitationally coupled fermions. From the point of view of the Standard Model sector these are electroweak singlets and so transform as would right-handed neutrinos. Although they are not forced to mix with Standard Model neutrinos, they are likely to do so unless this mixing is forbidden by a conservation law or selection rule (such as lepton number conservation). Because this neutrino mixing is renormalizable (and so is one of the portals) it is generically unsuppressed by super-heavy mass scales.

This makes a superpartner fermions look much like the light sterile neutrinos that are often postulated when neutrino model-building, though with an important extra ingredient: supersymmetry explains {\it why} they are light in the first place ({\it e.g.}~they might be superpartners for the massless graviton or for an extremely light dilaton). This is important because (unlike for perturbative Standard Model particles) chiral gauge interactions do not in themselves protect singlet fermions from getting very large masses. Indeed, some examples along these lines -- in which neutrinos mix with sterile fermions from the gravitational sector \cite{Dienes:1998sb} -- already exist in the literature.

\subsubsection*{Axions}

What was said above about fermions also applies to axions, since supersymmetric chiral multiplets involve complex scalars and this implies any light scalar field usually brings another along in its wake. These scalars are very often axions in the sense that they are the Goldstone bosons for shift symmetries and so tend to couple derivatively to ordinary matter (if at all). 

Derivative couplings and pseudoscalar parity make these scalars harder to detect, but also mean that they could well be ubiquitously present (as has been argued in \cite{axiverse} for instance). They also potentially give rise to many potentially detectable effects and are subject to a variety constraints \cite{Marsh:2015xka, Hook:2018dlk, Choi:2020rgn}.

A curious feature about axions that arise in supergravity is that they usually arise within UV completions as 2-form Kalb-Ramond gauge potentials, $b_{\mu\nu}$, that are related the the axion field $\mfa$ by a duality transformation like $\partial_\mu \mfa \propto \epsilon_{\mu\nu\lambda\rho} \partial^\nu \, b^{\lambda\rho}$. This means in particular that the corresponding axion need not combine with another field $\rho$ to form the combination $\rho \, e^{i\mfa}$ that linearly realizes its shift symmetry  at energies $E \gg f_a$, where $f_a$ is the axion's decay constant as defined by its kinetic term $f_a^2 (\partial \mfa)^2$. 

\subsubsection*{Primordial fluctuations and inflation}

Perhaps the most striking feature visible in expressions like \pref{Vsimp} or \pref{Vsimp2} for the scalar potential is the generic appearance of a large positive contribution to the scalar potential associated with the breaking scale of supersymmetry. This is precisely the kind of ingredient sought when building inflationary models, and one that is often hard to find when exploring UV completions (like string theory) because limitations in current calculational technology usually limit these to solutions are very close to a supersymmetric limit. 

Its ubiquity in the limit of strong supersymmetry breaking in the matter sector suggests that the scarcity of inflationary solutions in such searches is likely an artefact of the search techniques rather than being a robust consequence of UV physics. Indeed there is evidence \cite{StringX} that non-supersymmetric constructions (like physics localized on an antibrane within Type IIB string vacua) are well-described in the low-energy theory in terms of 4D supergravity coupled to a nilpotent goldstino field $X$.

Furthermore, if supersymmetric potentials can be relevant at the very low energies of late-time cosmology it is likely to be even more relevant at the higher energies at play during inflation. What the large positive supersymmetry breaking energies then suggest is that the energy scale of inflationary physics -- high though it is -- is likely lower than the scale associated with supersymmetry breaking itself. This also fits with the picture that more weakly coupled sectors appear to be more supersymmetric at lower energies because the shallowness of the inflaton potential usually requires its couplings to be quite small.

\subsubsection{Low-energy scalar potential}

Perhaps the most intriguing possibility that low-energy supersymmetry brings is the possibility for improved naturalness properties of the scalar potential at low energies. 

As mentioned above, the inevitable presence in the potential of the auxiliary fields associated with the supersymmetric gravity and goldstino sectors can change the nature of the scalar potential's UV sensitivity. For example, the contribution of a loop involving a dangerous particle of mass $M$ can contribute to the low-energy potential an amount $\delta V \sim M^2 F + \hbox{h.c.}$ (where $F$ is an auxiliary field) rather than the naive $\delta V \sim M^4$.  perhaps more generic light scalar fields? 

Although supersymmetry in itself is unlikely to make light scalars or small vacuum energies natural, it is likely to help other mechanisms for suppressing these quantities. (See in particular \cite{CCNo-Scale} for an example that attempts to combine supersymmetry and scale invariance to suppress UV contributions and so to obtain naturally light scalars and small vacuum energies.) 

Improved naturalness properties at very low energies would change much about the way we think about how fundamental physics can influence low-energy astronomy and cosmology, by removing the taboo on light cosmologically active dilaton-like scalars, possibly with many associated surprises. One such is the recent discovery that the presence of an axion with the couplings required to be the superpartner for a Brans-Dicke-like dilaton can make solar system tests unable to detect the dilaton even if its Brans-Dicke coupling is large enough that it would have been ruled out in the absence of the axion \cite{ADScreening}.

We welcome the possibilities of such a Brave New supersymmetric gravitational World!

\section*{Acknowledgements}
We thank Shanta de Alwis, Francesco Muia, Sergey Sibiryakov, Adam Solomon and Henry Tye for many helpful conversations.  CB's research was partially supported by funds from the Natural Sciences and Engineering Research Council (NSERC) of Canada. Research at the Perimeter Institute is supported in part by the Government of Canada through NSERC and by the Province of Ontario through MRI.  The work of FQ has been partially supported by STFC consolidated grants ST/P000681/1, ST/T000694/1.




\begin{thebibliography}{99}


\bibitem{ATLAS:2015wrn}
G.~Aad \textit{et al.} [ATLAS],
``Summary of the ATLAS experiment\textquoteright{}s sensitivity to supersymmetry after LHC Run 1 \textemdash{} interpreted in the phenomenological MSSM,''
JHEP \textbf{10} (2015), 134
[arXiv:1508.06608 [hep-ex]].

\bibitem{Sarkar:2021lju}
U.~Sarkar [CMS],
``Searches for supersymmetry in CMS,''
[arXiv:2105.01629 [hep-ex]].

\bibitem{ParticleDataGroup:2020ssz}
P.~A.~Zyla \textit{et al.} [Particle Data Group],
``Review of Particle Physics,''
PTEP \textbf{2020} (2020) no.8, 083C01

\bibitem{Green:2012oqa}
M.~B.~Green, J.~H.~Schwarz and E.~Witten,
``Superstring Theory Vol. 1: 25th Anniversary Edition,''
``Superstring Theory Vol. 2: 25th Anniversary Edition,'' Cambridge University Press (1986)

\bibitem{Polchinski:1998rr}
J.~Polchinski,
``String theory. Vol. 2: Superstring theory and beyond,'' Cambridge University Press (1998)

\bibitem{Becker:2006dvp}
K.~Becker, M.~Becker and J.~H.~Schwarz,
``String theory and M-theory: A modern introduction,'' Cambridge University Press (2006)

\bibitem{Ibanez:2012zz}
L.~E.~Ibanez and A.~M.~Uranga,
``String theory and particle physics: An introduction to string phenomenology,'' Cambridge University Press (2012).


\bibitem{Arkani-Hamed:2006emk}
N.~Arkani-Hamed, L.~Motl, A.~Nicolis and C.~Vafa,
``The String landscape, black holes and gravity as the weakest force,''
JHEP \textbf{06} (2007), 060
[arXiv:hep-th/0601001 [hep-th]].



\bibitem{SLED}
Y.~Aghababaie, C.~P.~Burgess, S.~L.~Parameswaran and F.~Quevedo,
``Towards a naturally small cosmological constant from branes in 6-D supergravity,''
Nucl. Phys. B \textbf{680} (2004), 389-414
[arXiv:hep-th/0304256 [hep-th]];

C.~P.~Burgess, J.~Matias and F.~Quevedo,
``MSLED: A Minimal supersymmetric large extra dimensions scenario,''
Nucl. Phys. B \textbf{706} (2005), 71-99
[arXiv:hep-ph/0404135 [hep-ph]];\\


\bibitem{HighEUV}
S.~B.~Giddings, S.~Kachru and J.~Polchinski,
``Hierarchies from fluxes in string compactifications,''
Phys. Rev. D \textbf{66} (2002), 106006
[arXiv:hep-th/0105097 [hep-th]];

S.~Kachru, R.~Kallosh, A.~D.~Linde and S.~P.~Trivedi,
``De Sitter vacua in string theory,''
Phys. Rev. D \textbf{68} (2003), 046005
[arXiv:hep-th/0301240 [hep-th]];

  V. Balasubramanian, P. Berglund, J. P. Conlon and F. Quevedo, 
  ``Systematics of moduli stabilisation in Calabi-Yau flux compactifications,'' 
  JHEP 0503 (2005) 007 [hep-th/0502058].


\bibitem{SUGRAQuint}
P.~Binetruy,
``Models of dynamical supersymmetry breaking and quintessence,''
Phys. Rev. D \textbf{60} (1999), 063502
[arXiv:hep-ph/9810553 [hep-ph]];

A.~Masiero, M.~Pietroni and F.~Rosati,
``SUSY QCD and quintessence,''
Phys. Rev. D \textbf{61} (2000), 023504
[arXiv:hep-ph/9905346 [hep-ph]];

P.~Brax and J.~Martin,
``Quintessence and supergravity,''
Phys. Lett. B \textbf{468} (1999), 40-45
[arXiv:astro-ph/9905040 [astro-ph]];

E.~J.~Copeland, N.~J.~Nunes and F.~Rosati,
``Quintessence models in supergravity,''
Phys. Rev. D \textbf{62} (2000), 123503
[arXiv:hep-ph/0005222 [hep-ph]].


\bibitem{Binetruy:2009zz}
P.~Binetruy,
``Dark energy and fundamental physics,''
J. Phys. Conf. Ser. \textbf{171} (2009), 012011

\bibitem{quintessence}
B.~Ratra and P.~J.~E.~Peebles,
``Cosmological Consequences of a Rolling Homogeneous Scalar Field,''
Phys. Rev. D \textbf{37} (1988), 3406;

R.~D.~Peccei, J.~Sola and C.~Wetterich,
``Adjusting the Cosmological Constant Dynamically: Cosmons and a New Force Weaker Than Gravity,''
Phys. Lett. B \textbf{195} (1987), 183-190;

C.~Wetterich,
``Cosmology and the Fate of Dilatation Symmetry,''
Nucl. Phys. B \textbf{302} (1988), 668-696
[arXiv:1711.03844 [hep-th]].

\bibitem{QuintessenceReviews}
P.~J.~E.~Peebles and B.~Ratra,
``The Cosmological Constant and Dark Energy,''
Rev. Mod. Phys. \textbf{75} (2003), 559-606
[arXiv:astro-ph/0207347 [astro-ph]];

E.~J.~Copeland, M.~Sami and S.~Tsujikawa,
``Dynamics of dark energy,''
Int. J. Mod. Phys. D \textbf{15} (2006), 1753-1936
[arXiv:hep-th/0603057 [hep-th]].


\bibitem{CCWeinberg}
S.~Weinberg,
``The Cosmological Constant Problem,''
Rev. Mod. Phys. \textbf{61} (1989), 1-23.

\bibitem{Burgess:2013ara}
C.~P.~Burgess,
``The Cosmological Constant Problem: Why it's hard to get Dark Energy from Micro-physics,''
[arXiv:1309.4133 [hep-th]].


\bibitem{Komargodski:2009rz}
Z.~Komargodski and N.~Seiberg,
``From Linear SUSY to Constrained Superfields,''
JHEP \textbf{09} (2009), 066
[arXiv:0907.2441 [hep-th]].


\bibitem{Bergshoeff:2015tra}
E.~A.~Bergshoeff, D.~Z.~Freedman, R.~Kallosh and A.~Van Proeyen,
``Pure de Sitter Supergravity,''
Phys. Rev. D \textbf{92} (2015) no.8, 085040
[erratum: Phys. Rev. D \textbf{93} (2016) no.6, 069901]
[arXiv:1507.08264 [hep-th]].

\bibitem{Dudas:2015eha}
E.~Dudas, S.~Ferrara, A.~Kehagias and A.~Sagnotti,
``Properties of Nilpotent Supergravity,''
JHEP \textbf{09} (2015), 217
[arXiv:1507.07842 [hep-th]].

\bibitem{DallAgata:2015zxp}
G.~Dall'Agata and F.~Farakos,
``Constrained superfields in Supergravity,''
JHEP \textbf{02} (2016), 101
[arXiv:1512.02158 [hep-th]].

\bibitem{Schillo:2015ssx}
M.~Schillo, E.~van der Woerd and T.~Wrase,
``The general de Sitter supergravity component action,''
Fortsch. Phys. \textbf{64} (2016), 292-302
[arXiv:1511.01542 [hep-th]].


\bibitem{CCNo-Scale}
C.P.~Burgess, D.~Dineen and F.~Quevedo,
``Dark Implications of a Supersymmetric Gravity Sector:
Scale Invariance and a Naturally Relaxed Dark Energy,''
[arXiv:2110.xxxxx [hep-th]].

\bibitem{Burgess:2020qsc}
C.~P.~Burgess, M.~Cicoli, D.~Ciupke, S.~Krippendorf and F.~Quevedo,
``UV Shadows in EFTs: Accidental Symmetries, Robustness and No-Scale Supergravity,''
Fortsch. Phys. \textbf{68} (2020) no.10, 2000076
[arXiv:2006.06694 [hep-th]].


\bibitem{QHE}
E.~Witten,
``Three lectures on topological phases of matter,''
Riv. Nuovo Cim. \textbf{39} (2016) no.7, 313-370
[arXiv:1510.07698 [cond-mat.mes-hall]]; 

D.~Tong,
``Lectures on the Quantum Hall Effect,''
[arXiv:1606.06687 [hep-th]],//

\bibitem{EFTBook}
 C.~P.~Burgess,
``Introduction to Effective Field Theory,''
Cambridge University Press (December 2020).


\bibitem{LuisForm}
S.~Bielleman, L.~E.~Ibanez and I.~Valenzuela,
``Minkowski 3-forms, Flux String Vacua, Axion Stability and Naturalness,''
JHEP \textbf{12} (2015), 119
[arXiv:1507.06793 [hep-th]]..


\bibitem{MyForm}
C.~P.~Burgess, R.~Diener and M.~Williams,
``Self-Tuning at Large (Distances): 4D Description of Runaway Dilaton Capture,''
JHEP \textbf{10} (2015), 177
[arXiv:1509.04209 [hep-th]].

\bibitem{Inflation}
S.~Ferrara, R.~Kallosh and A.~Linde,
``Cosmology with Nilpotent Superfields,''
JHEP \textbf{10} (2014), 143
[arXiv:1408.4096 [hep-th]];

S.~Ferrara, R.~Kallosh and J.~Thaler,
``Cosmology with orthogonal nilpotent superfields,''
Phys. Rev. D \textbf{93} (2016) no.4, 043516
[arXiv:1512.00545 [hep-th]].

\bibitem{axiverse}
A.~Arvanitaki, S.~Dimopoulos, S.~Dubovsky, N.~Kaloper and J.~March-Russell,
``String Axiverse,''
Phys. Rev. D \textbf{81} (2010), 123530
[arXiv:0905.4720 [hep-th]];

M.~Cicoli, M.~Goodsell and A.~Ringwald,
``The type IIB string axiverse and its low-energy phenomenology,''
JHEP \textbf{10} (2012), 146
[arXiv:1206.0819 [hep-th]].

\bibitem{Garny:2015sjg}
M.~Garny, M.~Sandora and M.~S.~Sloth,
``Planckian Interacting Massive Particles as Dark Matter,''
Phys. Rev. Lett. \textbf{116} (2016) no.10, 101302
[arXiv:1511.03278 [hep-ph]].

\bibitem{Dienes:1998sb}
K.~R.~Dienes, E.~Dudas and T.~Gherghetta,
``Neutrino oscillations without neutrino masses or heavy mass scales: A Higher dimensional seesaw mechanism,''
Nucl. Phys. B \textbf{557} (1999), 25
[arXiv:hep-ph/9811428 [hep-ph]];

J.~Matias and C.~P.~Burgess,
``MSLED, neutrino oscillations and the cosmological constant,''
JHEP \textbf{09} (2005), 052
[arXiv:hep-ph/0508156 [hep-ph]].

\bibitem{Weinberg:1968de}
S.~Weinberg,
``Nonlinear realizations of chiral symmetry,''
Phys. Rev. \textbf{166} (1968), 1568-1577

\bibitem{Coleman:1969sm}
S.~R.~Coleman, J.~Wess and B.~Zumino,
``Structure of phenomenological Lagrangians. 1.,''
Phys. Rev. \textbf{177} (1969), 2239-2247

\bibitem{Callan:1969sn}
C.~G.~Callan, Jr., S.~R.~Coleman, J.~Wess and B.~Zumino,
``Structure of phenomenological Lagrangians. 2.,''
Phys. Rev. \textbf{177} (1969), 2247-2250



\bibitem{VolkovAkulov}
D.~V.~Volkov and V.~P.~Akulov,
``Is the Neutrino a Goldstone Particle?,''
Phys. Lett. B \textbf{46} (1973), 109-110


\bibitem{West}
P.~West,
``Introduction to Supersymmetry and Supergravity,''
World Scientific (1986).

\bibitem{WB}
J.~Bagger and J.~Wess,
``Supersymmetry and supergravity,''
Princeton Series in Physics (1992)


\bibitem{FreedmanVanProeyen}
{\em Supergravity} by Daniel Z.~Freedman and Antoine Van Proeyen, Cambridge University Press (2012)

\bibitem{SuperHiggs}
E.~Cremmer, B.~Julia, J.~Scherk, P.~van Nieuwenhuizen, S.~Ferrara and L.~Girardello,
``Super-higgs effect in supergravity with general scalar interactions,''
Phys. Lett. B \textbf{79} (1978), 231-234;

E.~Cremmer,~S. Ferrara, L.~Girardello and A.~Van Proeyen, 
``Yang–Mills theories with local supersymmetry: Lagrangian, transformation laws and superhiggs effect,''
Nucl. Phys. \textbf{B212} (1983) 413;

M.~T.~Grisaru, M.~Rocek and A.~Karlhede, 
``The superhiggs effect in superspace,''
Phys. Lett. \textbf{B120} (1983) 110.
 
 \bibitem{DeWitt:1964mxt}
B.~S.~DeWitt,
``Dynamical theory of groups and fields,''
Conf. Proc. C \textbf{630701} (1964), 585-820

\bibitem{HeatKernel}
P.~B.~Gilkey,
``The Spectral Geometry Of A Riemannian Manifold,''
J.\ Diff.\ Geom.\  {\bf 10}, 601 (1975);

S.M.~Christensen, {\it Phys. Rev.} {\bf D17} (1978) 9460--963;

S.M.~Christensen and M.J.~Duff, {\it Nucl. Phys.} {\bf B154}
(1979) 301;

D.M.~McAvity and H.~Osborn, {\it Class. Quant. Grav.} {\bf 8}
(1991) 603--638.

\bibitem{HKReview}
For a  review of heat kernel techniques see:
D.~V.~Vassilevich, ``Heat kernel expansion: User's manual,''
Phys.\ Rep.\  {\bf 388}, 279 (2003) [arXiv:hep-th/0306138].

\bibitem{HKXD}
For heat kernel results for a variety of fields in $D$ dimensions see

D.~Hoover and C.~P.~Burgess,
``Ultraviolet sensitivity in higher dimensions,''
JHEP \textbf{01} (2006), 058
[arXiv:hep-th/0507293 [hep-th]];

C.~P.~Burgess and D.~Hoover,
``UV sensitivity in supersymmetric large extra dimensions: The Ricci-flat case,''
Nucl. Phys. B \textbf{772} (2007), 175-204
[arXiv:hep-th/0504004 [hep-th]].

\bibitem{ColemanWeinberg}
S.~R.~Coleman and E.~J.~Weinberg,
``Radiative Corrections as the Origin of Spontaneous Symmetry Breaking,''
Phys. Rev. D \textbf{7} (1973), 1888-1910;

S.~Weinberg,
``Perturbative Calculations of Symmetry Breaking,''
Phys. Rev. D \textbf{7} (1973), 2887-2910


\bibitem{SuperTrace}
S.~Ferrara, L.~Girardello and F.~Palumbo, 
``A general mass formula in broken supersymmetry,''
Phys. Rev. \textbf{D20} (1979) 403;

S.~Ferrara and A.~Van Proeyen,
``Mass Formulae for Broken Supersymmetry in Curved Space-Time,''
Fortsch. Phys. \textbf{64} (2016) no.11-12, 896-902
[arXiv:1609.08480 [hep-th]].

\bibitem{Ferrara:1994kg}
S.~Ferrara, C.~Kounnas and F.~Zwirner,
``Mass formulae and natural hierarchy in string effective supergravities,''
Nucl. Phys. B \textbf{429} (1994), 589-625
[erratum: Nucl. Phys. B \textbf{433} (1995), 255-255]
[arXiv:hep-th/9405188 [hep-th]].

\bibitem{DeAlwis:2021gou}
S.~P.~De Alwis,
``Wilsonian Effective Field Theory and String Theory,''
[arXiv:2103.13347 [hep-th]].

\bibitem{NRTheorems}
M.T.~Grisaru, W.~Siegel and M.~Rocek, 
``Improved Methods For Supergraphs,” 
Nucl. Phys. \textbf{B159} (1979) 429;

E.~Witten, 
``New Issues In Manifolds Of SU(3) Holonomy,” 
Nucl. Phys. \textbf{B268} (1986) 79;

M.~Dine, N. Seiberg, 
``Nonrenormalization Theorems in Superstring Theory,”
Phys. Rev. Lett. \textbf{57} (1986) 21;

C.P.~Burgess, A.~Font and F.~Quevedo, 
``Low-Energy Effective Action For The Superstring,” 
Nucl. Phys. \textbf{B272} (1986) 661;

N.~Seiberg, 
``Naturalness versus supersymmetric nonrenormalization theorems,” 
Phys. Lett. \textbf{B318} (1993) 469 
[hep-ph/9309335];

C.~P.~Burgess, C.~Escoda and F.~Quevedo,
``Nonrenormalization of flux superpotentials in string theory,''
JHEP \textbf{06} (2006), 044
[arXiv:hep-th/0510213 [hep-th]].

\bibitem{Antoniadis:2010hs}
I.~Antoniadis, E.~Dudas, D.~M.~Ghilencea and P.~Tziveloglou,
``Non-linear MSSM,''
Nucl. Phys. B \textbf{841} (2010), 157-177
[arXiv:1006.1662 [hep-ph]].

\bibitem{Li:2020rzo}
S.~Y.~Li, Y.~C.~Qiu and S.~H.~H.~Tye,
``Standard Model from A Supergravity Model with a Naturally Small Cosmological Constant,''
JHEP \textbf{05} (2021), 181
doi:10.1007/JHEP05(2021)181
[arXiv:2010.10089 [hep-th]].

\bibitem{Salam:1989fm}
  A.~Salam and E.~Sezgin,
  ``Supergravities In Diverse Dimensions. Vol. 1, 2,''
  Amsterdam, Netherlands: North-Holland (1989) 1499 p. Singapore, Singapore: World Scientific (1989) 1499 p
  
\bibitem{Burgess:2011rv}
  Y.~Aghababaie, C.~P.~Burgess, J.~M.~Cline, H.~Firouzjahi, S.~L.~Parameswaran, F.~Quevedo, G.~Tasinato and I.~Zavala,
  ``Warped brane worlds in six-dimensional supergravity,''
  JHEP {\bf 0309} (2003) 037
  [hep-th/0308064];
   
\bibitem{SUGRAscaleinv}
A.~J.~Tolley, C.~P.~Burgess, C.~de Rham and D.~Hoover,
``Scaling solutions to 6D gauged chiral supergravity,''
New J. Phys. \textbf{8} (2006), 324
[arXiv:hep-th/0608083 [hep-th]].
 
 \bibitem{BMvNNQ}
  C.~P.~Burgess, A.~Maharana, L.~van Nierop, A.~A.~Nizami and F.~Quevedo,
  ``On Brane Back-Reaction and de Sitter Solutions in Higher-Dimensional Supergravity,''
  JHEP {\bf 1204} (2012) 018
  [arXiv:1109.0532 [hep-th]];
  
\bibitem{GJZ}
  F.~F.~Gautason, D.~Junghans and M.~Zagermann,
  ``Cosmological Constant, Near Brane Behavior and Singularities,''
  JHEP {\bf 1309} (2013) 123
  [arXiv:1301.5647 [hep-th]].


\bibitem{Kawasaki:2008qe}
M.~Kawasaki, K.~Kohri, T.~Moroi and A.~Yotsuyanagi,
``Big-Bang Nucleosynthesis and Gravitino,''
Phys. Rev. D \textbf{78} (2008), 065011
[arXiv:0804.3745 [hep-ph]].

\bibitem{PDG}
P.A.~Zyla \textit{et al.} [Particle Data Group],
``Review of Particle Physics,''
PTEP \textbf{2020} (2020) no.8, 083C01


\bibitem{Feng:2010gw}
J.~L.~Feng,
``Dark Matter Candidates from Particle Physics and Methods of Detection,''
Ann. Rev. Astron. Astrophys. \textbf{48} (2010), 495-545
[arXiv:1003.0904 [astro-ph.CO]].

\bibitem{Coughlan:1983ci}
G.~D.~Coughlan, W.~Fischler, E.~W.~Kolb, S.~Raby and G.~G.~Ross,
``Cosmological Problems for the Polonyi Potential,''
Phys. Lett. B \textbf{131} (1983), 59-64

\bibitem{Banks:1993en}
T.~Banks, D.~B.~Kaplan and A.~E.~Nelson,
``Cosmological implications of dynamical supersymmetry breaking,''
Phys. Rev. D \textbf{49} (1994), 779-787
[arXiv:hep-ph/9308292 [hep-ph]].

\bibitem{deCarlos:1993wie}
B.~de Carlos, J.~A.~Casas, F.~Quevedo and E.~Roulet,
``Model independent properties and cosmological implications of the dilaton and moduli sectors of 4-d strings,''
Phys. Lett. B \textbf{318} (1993), 447-456
[arXiv:hep-ph/9308325 [hep-ph]].

\bibitem{Conlon:2007gk}
J.~P.~Conlon and F.~Quevedo,
``Astrophysical and cosmological implications of large volume string compactifications,''
JCAP \textbf{08} (2007), 019
[arXiv:0705.3460 [hep-ph]].


\bibitem{ScalarPortal}
M.~Veltman and F.~Yndurain, Nucl. Phys. B325 (1989) 1;

V.~Silveira and A.~Zee, Phys. Lett. B161 (1985) 136;

J.~McDonald,
``Gauge singlet scalars as cold dark matter,''
Phys. Rev. D \textbf{50} (1994), 3637-3649
[arXiv:hep-ph/0702143 [hep-ph]];

C.~P.~Burgess, M.~Pospelov and T.~ter Veldhuis,
``The Minimal model of nonbaryonic dark matter: A Singlet scalar,''
Nucl. Phys. B \textbf{619} (2001), 709-728
[arXiv:hep-ph/0011335 [hep-ph]].

\bibitem{Marsh:2015xka}
D.~J.~E.~Marsh,
``Axion Cosmology,''
Phys. Rept. \textbf{643} (2016), 1-79
[arXiv:1510.07633 [astro-ph.CO]].

\bibitem{Hook:2018dlk}
A.~Hook,
``TASI Lectures on the Strong CP Problem and Axions,''
PoS \textbf{TASI2018} (2019), 004
[arXiv:1812.02669 [hep-ph]].

\bibitem{Choi:2020rgn}
K.~Choi, S.~H.~Im and C.~S.~Shin,
``Recent progress in physics of axions or axion-like particles,''
[arXiv:2012.05029 [hep-ph]].


\bibitem{StringX}
R.~Kallosh and T.~Wrase,
``Emergence of Spontaneously Broken Supersymmetry on an Anti-D3-Brane in KKLT dS Vacua,''
JHEP \textbf{12} (2014), 117
[arXiv:1411.1121 [hep-th]];

I.~Antoniadis, E.~Dudas, S.~Ferrara and A.~Sagnotti,
``The Volkov\textendash{}Akulov\textendash{}Starobinsky supergravity,''
Phys. Lett. B \textbf{733} (2014), 32-35
[arXiv:1403.3269 [hep-th]];

L.~Aparicio, F.~Quevedo and R.~Valandro,
``Moduli Stabilisation with Nilpotent Goldstino: Vacuum Structure and SUSY Breaking,''
JHEP \textbf{03} (2016), 036
[arXiv:1511.08105 [hep-th]];

R.~Kallosh, F.~Quevedo and A.~M.~Uranga,
``String Theory Realizations of the Nilpotentz Goldstino,''
JHEP \textbf{12} (2015), 039
[arXiv:1507.07556 [hep-th]];

I.~Garc\'\i{}a-Etxebarria, F.~Quevedo and R.~Valandro,
``Global String Embeddings for the Nilpotent Goldstino,''
JHEP \textbf{02} (2016), 148
[arXiv:1512.06926 [hep-th]];

C.~Crin\`o, F.~Quevedo and R.~Valandro,
``On de Sitter String Vacua from Anti-D3-Branes in the Large Volume Scenario,''
JHEP \textbf{03} (2021), 258
[arXiv:2010.15903 [hep-th]].


\bibitem{ADScreening}
C.~P.~Burgess and F.~Quevedo,
``Axion Homeopathy: Screening Dilaton Interactions,''
[arXiv:2110.10352 [hep-th]].

\end{thebibliography}
\end{document}